\documentclass[a4paper,twoside,12pt]{article} 

\usepackage{a4wide}
\usepackage{graphicx}
\usepackage{amsmath,amssymb,amsthm,amsgen}
\usepackage{listings}
\usepackage{color}
\usepackage{hhline}
\usepackage{multirow}
\usepackage{enumitem}  
\usepackage{mathtools}
\usepackage[bookmarks]{}
\usepackage{booktabs}
\usepackage{amsfonts}   
\usepackage{mathrsfs}
\usepackage{bbm}
\usepackage{bm}  
\usepackage{tikz}
\usetikzlibrary{patterns, decorations.pathreplacing}
\usepackage{url}
\usepackage{stmaryrd}
\usepackage{longtable} 
\usepackage{framed}
\usepackage{thmtools,thm-restate} 
\usepackage{upgreek}
\usepackage{algorithm}
\usepackage{algorithmic}

\usepackage[hidelinks]{hyperref}
\long\def\comm#1{}

\theoremstyle{definition}

\newcommand{\E}{\mathop{{}\mathbb E}}

\newcommand{\N}{\mathbb N}
\newcommand{\R}{\mathbb R}

\newcommand{\T}{\mathbb T}

\newcommand{\W}{\mathbb W}

\newcommand{\Z}{\mathbb Z}
\def\softd{{\leavevmode\setbox1=\hbox{d}%
   \hbox to 1.05\wd1{d\kern-0.4ex{\char039}\hss}}}

\def\XXint#1#2#3{{\setbox0=\hbox{$#1{#2#3}{\int}$}
     \vcenter{\hbox{$#2#3$}}\kern-.5\wd0}}

\DeclareFontFamily{U}{mathx}{\hyphenchar\font45}
\DeclareFontShape{U}{mathx}{m}{n}{
      <5> <6> <7> <8> <9> <10>
      <10.95> <12> <14.4> <17.28> <20.74> <24.88>
      mathx10
      }{}
\DeclareSymbolFont{mathx}{U}{mathx}{m}{n}
\DeclareFontSubstitution{U}{mathx}{m}{n}
\DeclareMathAccent{\widecheck}{0}{mathx}{"71}

\newcommand\e{\varepsilon}

\newcommand{\bb}{\mathbf b}

\newcommand{\bL}{\mathbf L}

\long\def\drop#1{}

\newcommand\discann{(P^{\text{ann}}_n)}
\newcommand\disccon{(P^{\text{csv}}_n)}
\newcommand\contannNoParen{P^{\text{ann}}_\infty}
\newcommand\contconNoParen{P^{\text{csv}}_\infty}
\newcommand\contann{(\contannNoParen)}
\newcommand\contcon{(\contconNoParen)}
\newcommand\wann{ w^{\text{ann}} }
\newcommand\wcon{ w^{\text{csv}} }

\usepackage{ifthen}
\newlength{\leftstackrelawd}
\newlength{\leftstackrelbwd}
\def\leftstackrel#1#2{\settowidth{\leftstackrelawd}%
{${{}^{#1}}$}\settowidth{\leftstackrelbwd}{$#2$}%
\addtolength{\leftstackrelawd}{-\leftstackrelbwd}%
\leavevmode\ifthenelse{\lengthtest{\leftstackrelawd>0pt}}%
{\kern-.5\leftstackrelawd}{}\mathrel{\mathop{#2}\limits^{#1}}}

\begin{document}

\title{Simulations of dislocation dynamics on an atomic lattice: the effect of collision rules}

\author{Thomas Hudson \and Akaraphon Jantaraphum \and Patrick van Meurs}
\date{}

\maketitle


\begin{abstract}
The stochastic dynamics of dislocations on a one-dimensional periodic lattice domain are considered. Two models are studied: one without a collision rule, and one which annihilates colliding dislocations if they have opposite orientation. The behaviour of both models is investigated by means of a series of numerical simulations exploring the asymptotic behaviour of these models as the number of dislocations increases. From these simulations, evidence is obtained that the discrete model with annihilation tends to a PDE for the dislocation density that accounts for annihilation. However, the discrete model without a collision rule does not appear to exhibit consistent convergence behaviour; instead, it appears that the expected PDE with conserved dislocation density appears in the limit for some parameters, but that for other parameters the density appears to follow to the evolution of the PDE \textit{with annihilation}. These findings provide evidence that a careful treatment of dislocation collisions is important in discrete dislocation dynamics models. 
\end{abstract}


\clearpage

\tableofcontents

\section{Introduction}
\label{sec:intro}
The microscopic dynamics of interacting particle systems can often be modelled using macroscopic partial differential equations (PDEs). Theoretical scientists and mathematicians have long studied the connection between models at different scales, and there are a range of well-established theoretical tools to prove rigorously that particle densities converge to solutions of PDE models in appropriate asymptotic regimes. One aspect of an interacting particle system that can strongly affect the behaviour of asymptotic limits is the decay of interactions between particles; another is the possibility for microscopic interactions to have effects which are not easily captured at the PDE scale.

In this work, our focus is on models of interacting dislocations. Dislocations are line defects in the atomic lattice of metals. Their behaviour is an important factor in determining the plastic behaviour and strength of metals. They carry a vector-valued topological charge, called the Burgers vector, and their presence induces stress in the material that leads to non-local interactions. 

Due to the complexity of interacting curves and motivated by experimental observations in certain settings, it is often assumed that the dislocation lines are straight and parallel. This allows one to describe dislocations as particles on a two dimensional lattice perpendicular to the line direction. We will further assume that the Burgers vector of each dislocation is of equal length and parallel to the dislocation lines. Then, the Burgers vector is simply characterized by $+1$ or $-1$. The corresponding dislocations are known as screw dislocations.

The movement of dislocations to neighbouring lattice sites is driven by thermal activation. In line with other recent work in this area (see for example \cite{Hudson17,HudsonVanMeursPeletier20,Buze20}), we model this movement as a Markov process.  Modelling dislocation motion in this way presents both forms of challenging behaviour listed above: dislocations exert configurational forces on one another that decrease in size inversely with distance, and moreover their topological charge means that they may annihilate with one another when they collide. The combination of these two effects makes the analysis of these models very challenging. Advances have been made in recent research on the analysis of dislocation collisions in various settings 
\cite{ForcadelImbertMonneau09,
PatriziValdinoci15,
AlicandroDeLucaGarroniPonsiglione16,
VanMeursMorandotti19,
VanMeursPeletierPozar22,
PatriziSangsawang23,
VanMeursPatrizi24,
VanMeursPeletierSlangen25,
VanMeursTardyXX}, but rigorous connections from the atomistic scale to PDE models for the dislocation density remain out of reach.

Therefore, our core objective here is the pursuit of a \textit{numerical} investigation of these micro-to-macro connections. In particular, we investigate the effect of modelling choices concerning dislocation collisions. We find that different collision rules at the microscopic scale appear to lead to significantly different PDEs (or different interpretations of the same PDE system) at macroscopic scale. We obtain numerical evidence of both convergence and divergence between these models in a range of asymptotic regimes. 

The structure of the remainder of this work is as follows: in Section~\ref{sec:models}, we present the microscopic models of dislocation motion, and candidate PDE counterpart models at the macroscopic scale. In Section~\ref{sec:method} we outline the numerical methodology used to study both classes of problem. In Section~\ref{sec:results} we describe our simulations, and finally provide our conclusion on the convergence and divergence of the microscopic models in Section~\ref{sec:conclusion}.

\section{Models of dislocation motion}
\label{sec:models}

As mentioned above, the models considered in this work involve straight and parallel screw dislocations. 
Their Burgers vector is normalized to either $+1$ or $-1$. 
To keep the numerical computation time within workable limits, we assume that all dislocation lines move in the same plane, which makes the model effectively one dimensional. Within this one-dimensional setting, our model of screw dislocations is the same as that of edge dislocations.

\subsection{Discrete particle systems}
\label{sec:models:discrete}

\begin{figure}[b]
  \centering
  \begin{tikzpicture}[scale = 1.8]
   \def\r {.15}      
      
   \draw (-1.5,-.1) grid (6.5, .1);
   \draw (-1.5,0) node[left]{$\Lambda_\e$};

   \begin{scope}[shift={(2, \r + .1)}]   
     \filldraw[draw=red, thick, fill=white] (0, 0) circle (\r); 
     \draw[red, thick] (-.7*\r, 0) -- (.7*\r, 0);
     \draw[red, thick] (0, -.7*\r) -- (0, .7*\r);
   \end{scope}
   
   \begin{scope}[shift={(0, \r + .1)}]
     \filldraw[draw=blue, thick, fill=white] (0, 0) circle (\r); 
     \draw[blue, thick] (-.7*\r, 0) -- (.7*\r, 0);
   \end{scope}
   
   \begin{scope}[shift={(4, \r + .1)}]   
     \filldraw[draw=red, thick, fill=white] (0, 0) circle (\r); 
     \draw[red, thick] (-.7*\r, 0) -- (.7*\r, 0);
     \draw[red, thick] (0, -.7*\r) -- (0, .7*\r);
   \end{scope}
   
   \begin{scope}[shift={(5, \r + .1)}]
     \filldraw[draw=blue, thick, fill=white] (0, 0) circle (\r); 
     \draw[blue, thick] (-.7*\r, 0) -- (.7*\r, 0);
   \end{scope}
   
   \draw[<->] (-1,-.2) --++ (1,0) node[midway, below]{$\e$};
   
   \draw[->] (2 + \r, .1 + 2* \r) to [out=30,in=150, looseness=1] (3 - \r, .1 + 2*\r);
   \draw[->] (2 - \r, .1 + 2* \r) to [out=150,in=30, looseness=1] (1 + \r, .1 + 2*\r);
   
   \draw (2.5, .1 + 2.5*\r) node[above]{$r_{+,i}$};
   \draw (1.5, .1 + 2.5*\r) node[above]{$r_{-,i}$};
   \draw[red] (2,-.1) node[below]{$L_i$};
\end{tikzpicture}
  \caption{An illustration of the discrete particle model, showing part of the lattice $\Lambda_\e$ with several particles $L_j$ with signs $b_j$ on it (red is positive, blue is negative). The two possible jumps of $L_i$ are drawn together with the corresponding jump rates $r_{\pm, i}$. }
     \label{fig:discrete_model}
\end{figure}
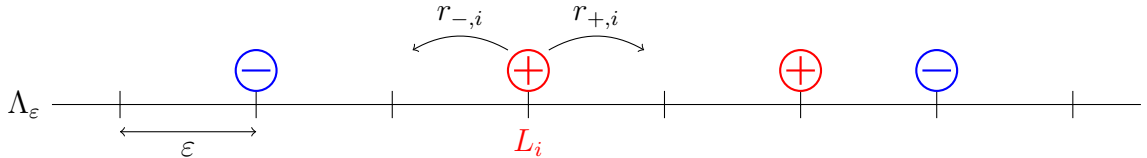

Here we describe in detail our two models $\disccon$ and $\discann$ for the discrete dislocation dynamics as continuous-time Markov chains. These models are small modification of that introduced in \cite{Hudson17}. Our two models only differ through the dislocation collision rule. The dynamics are illustrated in Figure~\ref{fig:discrete_model}. We first introduce the mathematical description in full, and comment on the underlying modelling choices afterwards.

Let 
\[
  \Lambda_\e := \e \Z / \Z = \{0,\e,2\e,\ldots,1-\e\}
\]
be the 1D periodic lattice with spacing  $\e > 0$ such that $\frac1\e \in \N$. We interpret $\Lambda_\e$ as the midpoints between neighboring atoms, where the dislocation cores are sited. $\e$ is the dimensionless ratio of the lattice spacing to the period normalized to $1$, which is the macroscopic length-scale.

On $\Lambda_\e$ we consider $n$ dislocations with position $L_i \in \Lambda_\e$ and normalized Burgers vector $b_i=\pm1$. We consider $\bb := (b_1, \ldots, b_n)$ as data and the dislocation configuration
\[
  \bL = (L_1, \ldots, L_n) \in (\Lambda_\e)^n
\]
as the state of the Markov chain.

To describe the Markov chain, it is sufficient to describe the jump rates between any pair of different states in $(\Lambda_\e)^n$. We set almost all rates to $0$, allowing nonzero rates only between those states that are obtained from one other by moving one dislocation $L_i$ to one of the two neighboring sites $L_i \pm \e \in \Lambda_\e$ (periodicity entails that the sites $0, 1-\e \in \Lambda_\e$ are neighbours). The formulae for the corresponding jump rates are
\begin{equation} \label{ri}
  r_{\pm,i}(\bL) := \dfrac{1}{\beta \e^2} \exp{\left(\pm \dfrac{1}{2} \beta \e F_i(\mathbf{L})\right)} > 0,
\end{equation}
where $\beta > 0$ is a dimensionless parameter that represents the ratio between the characteristic interaction energy scale and the characteristic thermal energy scale, and 
\begin{equation} \label{Fi}
  F_i: \Lambda_{\e}^n \to \R^n, \qquad F_i(\bL) := \dfrac{1}{n} \sum_{j=1}^n b_i b_j f(L_i-L_j)
\end{equation}
is the `force' acting on dislocation $i$. The product $b_i b_j$ determines the sign; it is positive for dislocations with the same sign (i.e.\ the same Burgers vector) and negative otherwise. The function
\begin{equation} \label{f:simple}
  f: \T \to \R, \qquad 
  f(x) := \begin{cases}
    \displaystyle \frac\pi{\tan \pi x} = - \pi \tan \Big( \pi \Big( x - \tfrac12 \Big) \Big), & x\notin\Z\\[2mm]
    0 & x\in\Z.
  \end{cases}
\end{equation}
is the dimensionless form of the dislocation interaction force. Note that $f(x) = \frac1x + O(x)$ for $|x| \ll 1$. The sign is such that dislocations of the same sign repel each other.
This completes the description of our first Markov chain model, which we refer to as $\disccon$. 
  
As the Markov process $(\bL(t))_{t \ge 0}$ evolves, we may encounter situations in which a dislocation hops to a site occupied by another dislocation. We call this situation a \textit{dislocation collision}. The choice on how to resolve this collision dictates the difference between the models $\disccon$ and $\discann$ for $(\bL(t))_{t \ge 0}$.
First, the model $\disccon$ does not treat collisions as special events. Note that our choice $f(0) = 0$ implies that the colliding dislocations do not interact with each other. Second, the model $\discann$ annihilates (i.e., removes from the system) instantaneously colliding dislocations if their Burgers vectors are of opposite sign. Technically, this can be described mathematically by expanding the state space of the Markov chain such that the length of $\bL$ is variable. We omit a full description, noting only that we keep the prefactor $\frac1n$ in \eqref{Fi} unaltered (where $n$ is the initial number of particles) as the system evolves. Note that while the number of dislocations will decrease over time, the \textit{net Burgers vector} $\sum_i b_i$ (with the sum running over the indices corresponding to the surviving particles) is conserved. Clearly, in $\disccon$ the dislocations are all preserved even after collision. As such, both the net Burgers vector $\sum_i b_i$ and the absolute Burgers vector $\sum_i |b_i| = n$ are conserved, motivating the label `$\disccon$'.

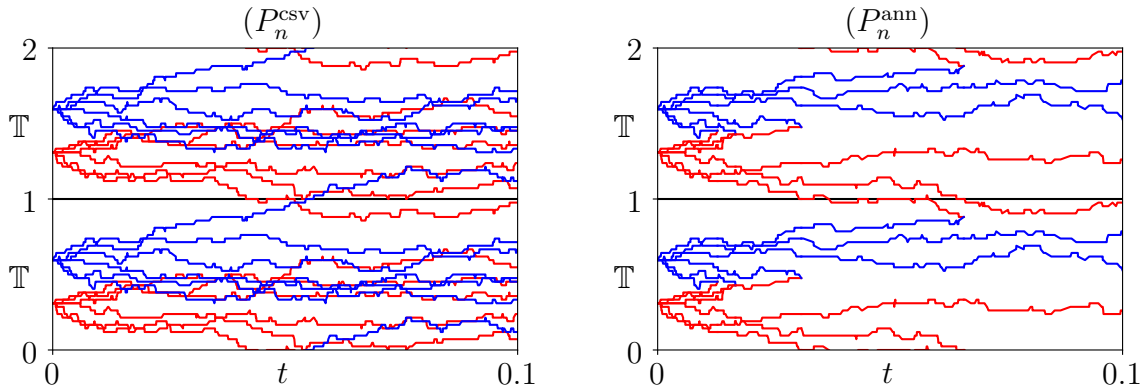
\begin{figure}[h!]
  \centering
  \begin{tikzpicture}
   \def\r {.15}      
      
   \draw (0,0) node{\includegraphics[width = .4\textwidth]{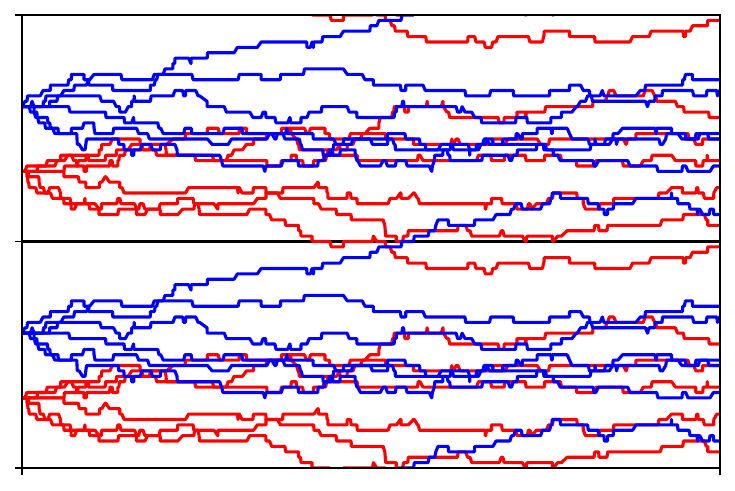}};
   \draw (-3.05,-2) node[below]{$0$};
   \draw (0,-2) node[below]{$t$};
   \draw (0,2) node[above]{$\disccon$};
   \draw (3.1,-2) node[below]{$0.1$};
   \draw (-3.1,-2) node[left]{$0$};
   \draw (-3.2,-1) node[left]{$\T$};
   \draw (-3.1,0) node[left]{$1$};
   \draw (-3.2,1) node[left]{$\T$};
   \draw (-3.1,2) node[left]{$2$};

   \begin{scope}[shift = {(8, 0)}]  
     \draw (0,0) node{\includegraphics[width = .4\textwidth]{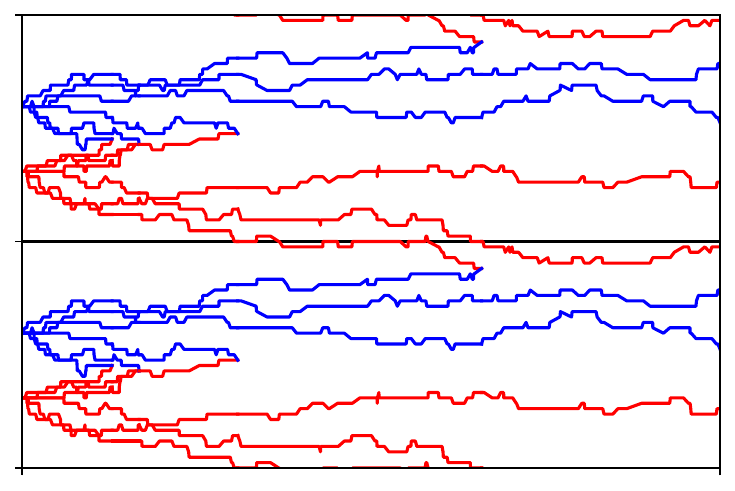}};
     
     \draw (-3.05,-2) node[below]{$0$};
   \draw (0,-2) node[below]{$t$};
   \draw (0,2) node[above]{$\discann$};
   \draw (3.1,-2) node[below]{$0.1$};
   \draw (-3.1,-2) node[left]{$0$};
   \draw (-3.2,-1) node[left]{$\T$};
   \draw (-3.1,0) node[left]{$1$};
   \draw (-3.2,1) node[left]{$\T$};
   \draw (-3.1,2) node[left]{$2$};      
   \end{scope}
\end{tikzpicture}
  \caption{Sample paths for the two discrete dislocation motion models. Dislocation positions are plotted against time $t$ over two spatial periods. The color indicates $b_i$; red is positive and blue is negative. The simulation parameters are $n=12$, $\e = 1/[n^{1.5}]$, $\beta = n^{0.8}$ with an identical random seed.}
     \label{fig:trajs}
\end{figure}

Example trajectories for the two models $\disccon$ and $\discann$ are shown in Figure~\ref{fig:trajs}. The differences between the models are apparent even though one is a modification of the other. The initial dispersion is caused by both the stochasticity and the repulsive force between same-sign dislocations. The effect of the attraction between opposite-sign dislocations is not so apparent (which might be because of the periodicity), but it is clear that paths of opposite color cross much more often than paths of the same color.

\paragraph{Underlying modelling choices.}
The motivations for our models are largely the same as in \cite{Hudson17,Buze20}, and are inspired more generally by approaches widely applied in computational materials science \cite{Voter07}. We briefly recall the motivation from those works. Suppose the collection of dislocations at sites $\bL$ in $\Lambda_\e$ provide a descriptor for local energy wells of the full atomistic energy. For the atomistic system to move to the neighbourhood of a new local minimum where a single dislocation has moved to an adjacent site, an energy barrier $\Delta E$ has to be overcome. To compute the rate in time at which this barrier is overcome, we assume low temperature such that Kramers' law can be applied \cite{Kramers1940,HTB90}. In particular, it states that the jump rate takes the form
\begin{equation*}
  r = \nu \exp(-\beta \Delta E). 
\end{equation*}
where $\nu > 0$ is an entropic prefactor. We have no direct physical motivation for the choice of $\nu$ being a uniform constant and of the precise form chosen in \eqref{ri}; instead, we follow the non--dimensionalisation argument conducted in \cite{HudsonVanMeursPeletier20}.

A further advantage of such models over direct atomistic simulation is that trajectories of such Markov chain models are easy and relatively inexpensive to compute numerically. We will apply the Kinetic Monte Carlo method, which is used across computational materials science to predict a range of thermally-activated metastable processes \cite{Voter07}.

Our choice for the form of the energy barrier $\Delta E$ comes from the approximation of the atomistic energy by classical linear elasticity theory \cite{Hudson17,Buze20}. 
Then, $F_i$ turns into a sum over the dislocation interactions, which are given by Volterra's formula $f$ for the Peach--Koehler force acting on dislocations. In the periodic setting, $f$ is alternatively obtained as the derivative of the continuum approximation of the lattice Green's function, which describes the atomistic interaction energy between screw dislocations. 

Finally, regarding the collision rule, $\discann$ is closest to the actual behaviour of dislocations. Indeed, the topological defects represented by dislocations with opposite Burgers vector cancel each other out upon collision, and `heal' the lattice locally.

\subsection{Parameter choices}
\label{sec:models:parameters} 

The key parameters which determine the dislocation kinetics in the discrete systems $\disccon$ and $\discann$ are:
\begin{itemize}\itemsep0em
\item $\e$, the ratio of the lattice spacing to macroscopic period;
\item $n$, the number of dislocations; and
\item $\beta$, the ratio between the interaction energy and the thermal energy scales.
\end{itemize}
In order to bridge the microscopic models $\disccon$ and $\discann$ to macroscopic models for dislocation densities, we have to make a choice for the asymptotic regime in the parameter space $(\e, n, \beta)$. We consider the regime in which:
\begin{itemize}
  \item $\frac1\e, n \gg 1$. A sequence of models where $n$ and $\frac{1}{\e}$ grow represents an ever-growing system size, for which we consider the asymptotic behaviour as the macroscopic dynamics.
  \item $n \ll \frac1\e$. This represents a dilute scaling regime for dislocations. The weaker condition $n \leq \frac1\e$ would guarantee that $\Lambda_\e$ has more sites than the number of dislocations, which we  deem as a necessary modelling assumption. Our stronger assumption pushes this further; $n \ll \frac1\e$ entails that the number of lattice sites grows asymptotically faster than the number of dislocations. This reflects that in most materials, the mean typical distance between dislocations is much greater than the lattice spacing. 
  \item $\beta \to \infty$. This is the low temperature regime, in which the thermal energy available for dislocation motion is much smaller than the elastic interaction energy. We expect diffusive effects to be asymptotically small in this setting.  
This is a common modelling choice in dislocation modelling, where thermal fluctuations are not considered directly; see for example the discussion in \cite[Chapter 10]{BulatovCai06}.
  \item $\beta \ll \frac1\e$. 
  This is a restriction on the physically interesting range. This choice is based in part on the results of \cite{BonaschiCarrilloDi-FrancescoPeletier15,BP16,Hudson17} for a simple choice of $F_i$. Here we motivate it formally. With this parameter choice, we can expand \eqref{ri} as
\[
  r_{\pm,i}(\bL) = \dfrac{1}{\beta \e^2} \left(1 \pm \dfrac{1}{2} \beta \e F_i(\mathbf{L})  + O(\beta^2 \e^2) \right).
\]
The leading order term leads to a random walk of $L_i$ on $\Lambda_\e$, while the next term describes a preferred direction. This demonstrates the result of the modelling choice $\beta \ll \frac1\e$; the randomness dominates the interaction force on the dislocation dynamics on the atomistic scale. The simulation in Figure~\ref{fig:trajs} illustrates this. 

From this modelling choice  we expect from the analysis in \cite{BonaschiCarrilloDi-FrancescoPeletier15} that the limiting equation for the dislocation density is a continuity equation. 
Instead, if $\beta \gg \frac1\e$, then one would instead expect rate-independent evolution for the dislocation density. Investigating the latter is beyond the scope of the present work.
\end{itemize}

\subsection{PDE models}
\label{sec:models:PDE}
The macroscopic counterparts of our discrete stochastic models $\disccon$ and $\discann$ are postulated mean-field limits $\contcon$ and $\contann$ respectively. They are PDEs that govern the evolution of the dislocation density over the 1D torus $\T \cong [0,1)$. 

In more detail, $\T$ is the continuous counterpart of the discrete lattice $\Lambda_\e$. The dislocation density functions are $\rho^+,\rho^-:[0,T) \times \T\to[0,\infty)$; one for the positive and one for the negative dislocations.  $T > 0$ is a given end time. Both densities are normalized such that, at least initially, $\int_\T \rho^\pm(t,x) \, dx = \frac{n^\pm}n$, where $n^\pm$ are the initial number of positive or negative dislocations. For convenience, we also introduce the net Burgers vector density $\kappa$, and the absolute dislocation density $\rho$ where
\[
\kappa := \rho^+-\rho^-\quad\text{and}\quad\rho:=\rho^++\rho^-.
\] 

Based on the analysis in \cite{GarroniVanMeursPeletierScardia19,HudsonVanMeursPeletier20} on models related to $\disccon$, we pick $\contcon$ as the Groma-Balogh equations \cite{GromaBalogh99}. These equations are formally given by
\begin{equation} \label{Pcon} \tag{$\contconNoParen$}
  \left\{ \begin{aligned} 
    \partial_t \rho^+ 
    &= - \partial_x (\rho^+ v[\kappa])
    &&\text{on } (0,T) \times \T  \\
    \partial_t \rho^- 
    &= + \partial_x (\rho^- v[\kappa])
    &&\text{on } (0,T) \times \T, 
  \end{aligned} \right.
  \qquad v[\kappa] := f * \kappa,
\end{equation}
where $v$ is the velocity (or ambient stress field), $f$ is the interaction force from \eqref{f:simple}, and $*$ indicates the convolution given by
\[
  (f * \kappa)(t,x) := \int_0^1 f(x-y) \kappa(t,y) \, dy.
\]
Since $f$ is singular at $0$, we interpret this definition as a principal value integral. We note that the only difference between the equations for $\rho^+$ and $\rho^-$ is the sign in front of the right-hand side. This difference reflects the effect that dislocations of opposite sign move in opposite direction under the same ambient shear stress.

Regarding $\discann$, we are not aware of a macroscopic system for which the limiting equation is known. Instead, a loosely related system was studied in \cite{VanMeursPeletierPozar22}. This involves signed dislocations with the same collision rule, but the dynamics is posed as a deterministic ODE system on $\R$ (instead of the periodic lattice $\Lambda_\e$). There, the limiting macroscopic equation was proven to be as follows, but instead posed on $\R$:
\begin{equation} \label{Pann} \tag{$\contannNoParen$} 
  \partial_t \kappa = - \partial_x (|\kappa| v[\kappa])
  \qquad \text{on } (0,T) \times \T,
\end{equation}
where $\kappa$ is the net Burgers vector density introduced above. In this model, the positive and negative densities of dislocations are recoverable by considering the positive and negative parts defined by
\begin{equation} \label{posneg:part}
  [\kappa]_+(x) := \begin{cases}
    \kappa(x) & \kappa(x)\geq0\\
    0 & \text{otherwise},
  \end{cases}\qquad\text{and}\qquad
  [\kappa]_-(x) := \begin{cases}
    -\kappa(x) & \kappa(x)\leq0\\
    0 & \text{otherwise.}
  \end{cases}
\end{equation}
We note that the model \eqref{Pann} goes back at least to \cite{Head72III}. It is similar to \eqref{Pcon} in the following sense: By taking the difference between the two equations of \eqref{Pcon}, we almost obtain \eqref{Pann}; the only difference is that $|\kappa|$ is replaced with $\rho$. At first sight it is not obvious whether this difference matters. In fact, on parts of $\T$ where either $\rho^+ = 0$ or $\rho^- = 0$, we have $|\kappa| = \rho$. In particular, for single-signed initial conditions, the equations are equivalent.

In what follows, we will reserve the symbols $\rho^\pm$ for solutions of \eqref{Pcon} and $\kappa$ for solutions of \eqref{Pann}. With this convention, it is clear by which model the two net Burgers vector densities $\rho^+ - \rho^-$ and $\kappa$ are computed. Similarly, we distinguish the positive and negative dislocation densities as $(\rho^+, \rho^-)$ and $([\kappa]_+, [\kappa]_-)$.

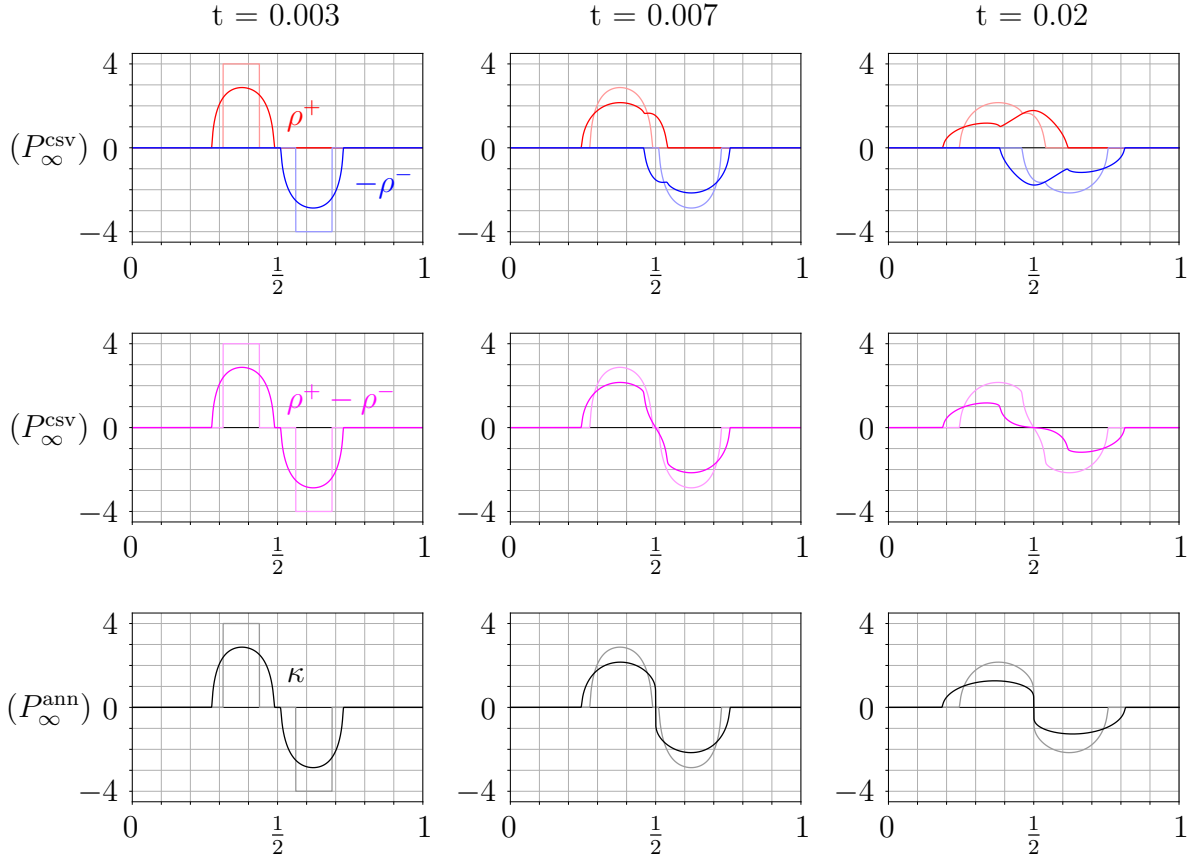
\begin{figure}[h!] 
  \centering
   \begin{tikzpicture}
    \def\r {.15}         
    \definecolor{mag}{RGB}{255,0,255}
    
    \draw (0,1.5) node[above] {t = 0.003};
    \draw (5,1.5) node[above] {t = 0.007};
    \draw (10,1.5) node[above] {t = 0.02};
    \draw (-3,0) node {$\contcon$};
    \draw (-3,-3.7) node {$\contcon$};
    \draw (-3,-7.4) node {$\contann$};
    
    \begin{scope}[shift={(0,0)}, scale = .4417]
       \draw (0,0) node{\includegraphics[width = .25\textwidth]{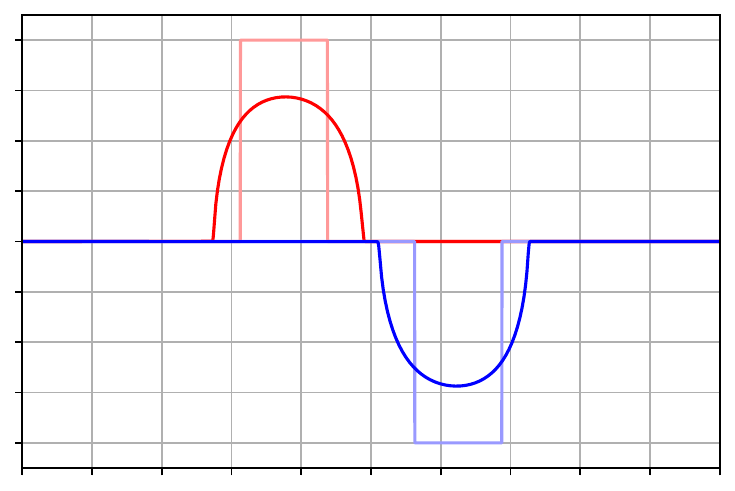}};
       \draw (-4.35,-2.9) node[below]{$0$};
       \draw (0.05,-2.9) node[below]{$\frac12$};
       \draw (4.45,-2.9) node[below]{$1$};
       \draw (-4.4,-2.5) node[left]{$-4$};
       \draw (-4.4,0) node[left]{$0$};
       \draw (-4.4,2.55) node[left]{$4$};
       \draw[red] (0,1) node[right] {$\rho^+$};
       \draw[blue] (2,-1) node[right] {$-\rho^-$};
    \end{scope}

    \begin{scope}[shift={(5,0)}, scale = .4417]
       \draw (0,0) node{\includegraphics[width = .25\textwidth]{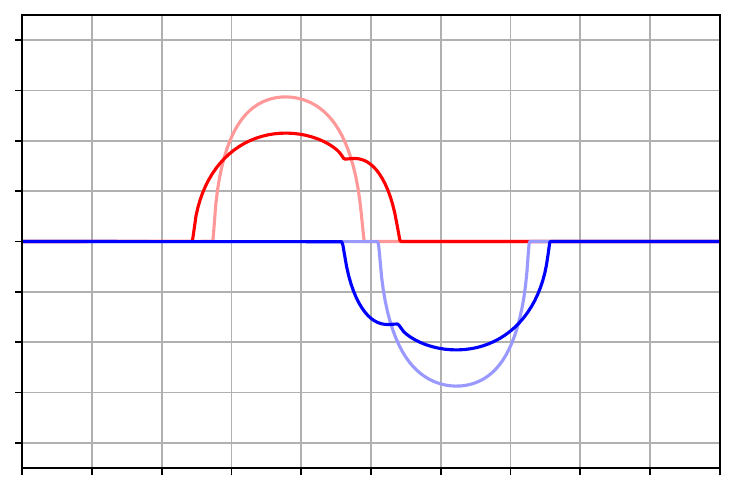}};
       \draw (-4.35,-2.9) node[below]{$0$};
       \draw (0.05,-2.9) node[below]{$\frac12$};
       \draw (4.45,-2.9) node[below]{$1$};
       \draw (-4.4,-2.5) node[left]{$-4$};
       \draw (-4.4,0) node[left]{$0$};
       \draw (-4.4,2.55) node[left]{$4$};
    \end{scope}
    
    \begin{scope}[shift={(10,0)}, scale = .4417]
       \draw (0,0) node{\includegraphics[width = .25\textwidth]{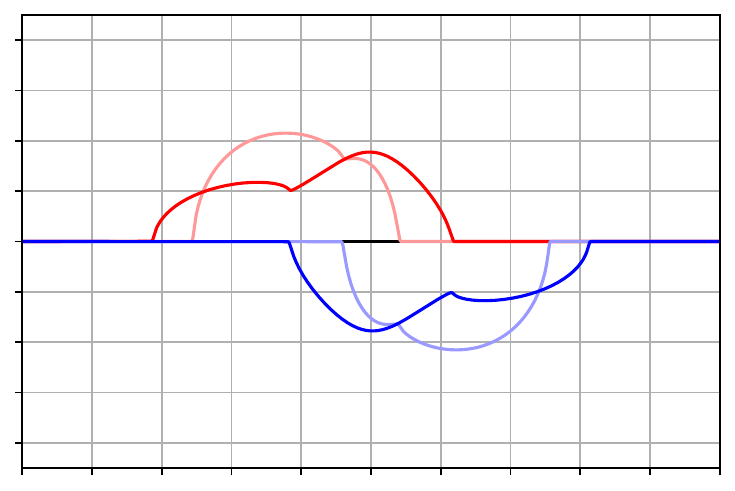}};
       \draw (-4.35,-2.9) node[below]{$0$};
       \draw (0.05,-2.9) node[below]{$\frac12$};
       \draw (4.45,-2.9) node[below]{$1$};
       \draw (-4.4,-2.5) node[left]{$-4$};
       \draw (-4.4,0) node[left]{$0$};
       \draw (-4.4,2.55) node[left]{$4$};
    \end{scope}
    
    \begin{scope}[shift={(0,-3.7)}, scale = .4417]
       \draw (0,0) node{\includegraphics[width = .25\textwidth]{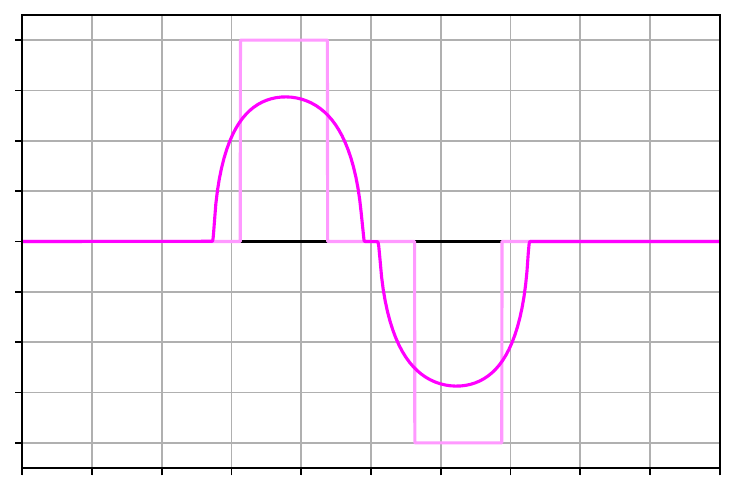}};
       \draw (-4.35,-2.9) node[below]{$0$};
       \draw (0.05,-2.9) node[below]{$\frac12$};
       \draw (4.45,-2.9) node[below]{$1$};
       \draw (-4.4,-2.5) node[left]{$-4$};
       \draw (-4.4,0) node[left]{$0$};
       \draw (-4.4,2.55) node[left]{$4$};
       \draw[color=mag] (0,.9) node[right] {$\rho^+ - \rho^-$};
    \end{scope}
    
    \begin{scope}[shift={(5,-3.7)}, scale = .4417]
       \draw (0,0) node{\includegraphics[width = .25\textwidth]{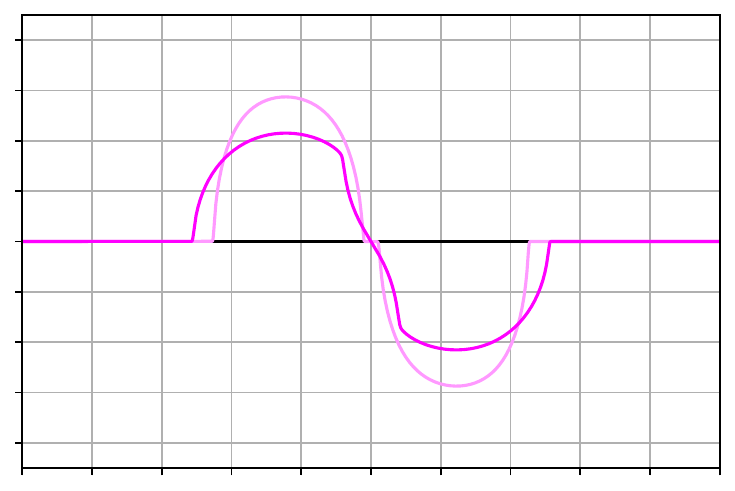}};
       \draw (-4.35,-2.9) node[below]{$0$};
       \draw (0.05,-2.9) node[below]{$\frac12$};
       \draw (4.45,-2.9) node[below]{$1$};
       \draw (-4.4,-2.5) node[left]{$-4$};
       \draw (-4.4,0) node[left]{$0$};
       \draw (-4.4,2.55) node[left]{$4$};
    \end{scope}
    
    \begin{scope}[shift={(10,-3.7)}, scale = .4417]
       \draw (0,0) node{\includegraphics[width = .25\textwidth]{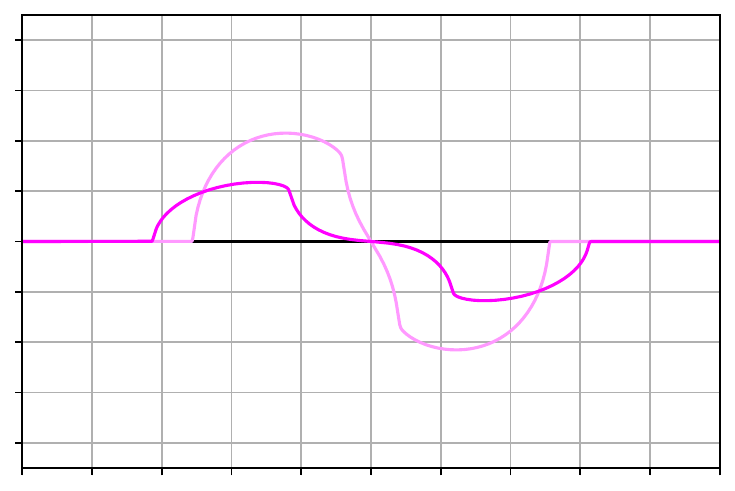}};
       \draw (-4.35,-2.9) node[below]{$0$};
       \draw (0.05,-2.9) node[below]{$\frac12$};
       \draw (4.45,-2.9) node[below]{$1$};
       \draw (-4.4,-2.5) node[left]{$-4$};
       \draw (-4.4,0) node[left]{$0$};
       \draw (-4.4,2.55) node[left]{$4$};
    \end{scope}
    
    \begin{scope}[shift={(0,-7.4)}, scale = .4417]
       \draw (0,0) node{\includegraphics[width = .25\textwidth]{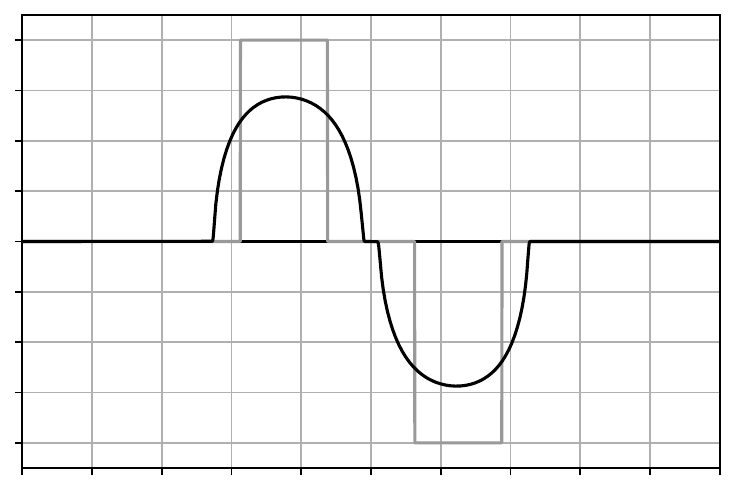}};
       \draw (-4.35,-2.9) node[below]{$0$};
       \draw (0.05,-2.9) node[below]{$\frac12$};
       \draw (4.45,-2.9) node[below]{$1$};
       \draw (-4.4,-2.5) node[left]{$-4$};
       \draw (-4.4,0) node[left]{$0$};
       \draw (-4.4,2.55) node[left]{$4$};
       \draw (0,1) node[right] {$\kappa$};
    \end{scope}
    
    \begin{scope}[shift={(5,-7.4)}, scale = .4417]
       \draw (0,0) node{\includegraphics[width = .25\textwidth]{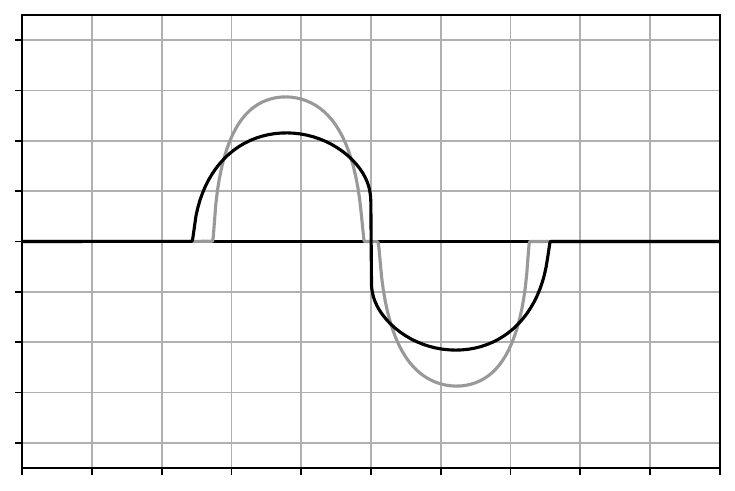}};
       \draw (-4.35,-2.9) node[below]{$0$};
       \draw (0.05,-2.9) node[below]{$\frac12$};
       \draw (4.45,-2.9) node[below]{$1$};
       \draw (-4.4,-2.5) node[left]{$-4$};
       \draw (-4.4,0) node[left]{$0$};
       \draw (-4.4,2.55) node[left]{$4$};
    \end{scope}
    
    \begin{scope}[shift={(10,-7.4)}, scale = .4417]
       \draw (0,0) node{\includegraphics[width = .25\textwidth]{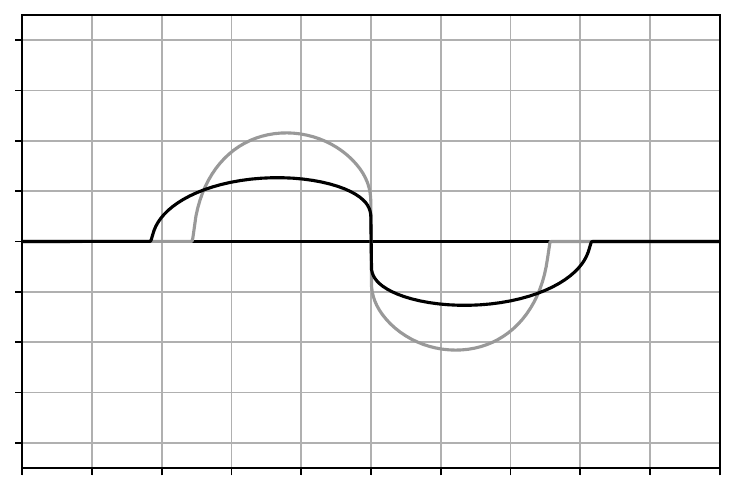}};
       \draw (-4.35,-2.9) node[below]{$0$};
       \draw (0.05,-2.9) node[below]{$\frac12$};
       \draw (4.45,-2.9) node[below]{$1$};
       \draw (-4.4,-2.5) node[left]{$-4$};
       \draw (-4.4,0) node[left]{$0$};
       \draw (-4.4,2.55) node[left]{$4$};
    \end{scope}
 \end{tikzpicture} 
    \caption{The density profiles for the solutions of the continuum models \eqref{Pcon} and \eqref{Pann} at three different times, starting from equivalent initial conditions (given by the faint graphs in the left column). The first row shows individual densities $\rho^+$ and $\rho^-$ evolving under $\contcon$; the second row shows the signed density $\rho^+-\rho^-$ predicted by the same model. The third row shows the evolution of $\kappa$ under $\contann$. We see stark differences between the models after the positive and negative dislocation densities come into contact. In each subfigure, the faint graphs show the profiles at the previous time for comparison.
    }
    \label{fig:rho:kap:intro}
\end{figure}

Next we examine the qualitative behaviour of the solutions to $\contcon$ and $\contann$. Figure~\ref{fig:rho:kap:intro} illustrates the solutions for a simple choice of the initial conditions where we take two spatially-separated, compactly-supported densities of postive and negative dislocations with the same block-shape.  By symmetry, i.e.\ $\rho^-(t,x) = \rho^+(t,\frac12-x)$, it is sufficient to describe the evolution of $\rho^+$. Initially, the profile of $\rho^+$ spreads out through self-repulsion, and its centre of mass moves slightly towards that of $\rho^-$ due to elastic attraction. The profile is approximately oval, in line with the analysis of \cite{BilerKarchMonneau10}, where it is shown that self-similar solutions of \eqref{Pann} posed on $\R$ are half-ellipses. As the evolution progresses beyond the point where $\rho^+$ and $\rho^-$ touch, a second bubble shape appears to be formed in the profile of $\rho^+$ over the region where $\rho^+$ and $\rho^-$ are both supported. In time, this second bubble seems to grow in size whereas the original bubble decreases in size. The overlapping, growing bubbles could be interpreted as annihilation, at least from the point of view of an observer: the signed density prediction on the second row demonstrate this. Indeed, it shows that $\int_\T |\rho^+ - \rho^-| dx$ is decreasing in time.

The evolution of the solution $\kappa$ of \eqref{Pann} shares some similar features with that of $\contcon$, but there are also clear differences when compared to the signed density $\rho^+ - \rho^-$ derived from \eqref{Pcon}. The main difference is that a second bubble is not formed. Instead, $\kappa$ has a steep interface at its front where it changes sign, whereas $\rho^+ - \rho^-$ has a gradual, smooth transition, due to the different handling of the annihilated dislocation mass by the PDEs. A possible explanation for the occurrence of a steep interface is that the PDE is of conservation-law type, and thus shocks may form. 

We now speculate on some basic properties of the models \eqref{Pcon} and \eqref{Pann}. The equations of \eqref{Pcon} are continuity equations, and thus $\int_\T \rho^\pm\,\mathrm{d}x$ are conserved. \eqref{Pann} is not a continuity equation, but from the nature of the solution concept proposed in \cite{VanMeursPeletierPozar22} for a similar system it appears that $\int_\T\kappa\mathrm{d}x$ is conserved. Our numerical schemes (see Section \ref{sec:FVdiscretisation}) are constructed to conserve these properties.

Finally, we discuss the well-posedness of \eqref{Pcon} and \eqref{Pann}. Existence of weak solutions for the PDE system \eqref{Pcon} was proved in 
\cite{MasmoudiZhang05,GarroniVanMeursPeletierScardia19}. In particular, it was shown in \cite{GarroniVanMeursPeletierScardia19} that under appropriate assumptions on the initial condition for the densities, the solutions satisfy $\rho^\pm(t,\cdot)\in L^1(\T)$ and $v[\kappa(t,\cdot)] \in H^{1/2}(\T)$ for almost every time $t$. We are not aware of any uniqueness results for $\contcon$. Regarding \eqref{Pann}, when posed on $\R$, the existence and uniqueness of strong solutions was proven in \cite{MasmoudiZhang05}, and the existence and uniqueness of viscosity solutions in terms of the cumulative density $u(x) := \int_{-\infty}^x \kappa$ was proven in \cite{BilerKarchMonneau10} with a crucial construction coming from \cite{ImbertMonneauRouy08}. From this we conjecture the well-posedness of \eqref{Pann} on $\T$, but proving such a result is beyond the scope of the current work.

\section{Computational methodology}
\label{sec:method}

In this section we explain how we numerically approximate solutions of $\disccon$, $\discann$, \eqref{Pcon} and \eqref{Pann}. We also demonstrate how we choose consistent initial conditions, and how we compare the solutions of the Markov chains to the solutions of the PDEs. Finally, we numerically investigate convergence of the models, i.e. whether `$\disccon \to \contcon$' and `$\disccon \to \contcon$' as $n \to \infty$ in an appropriate sense.

\subsection{Kinetic Monte Carlo approach}
\label{sec:method:KMC}

To simulate samples of the solution trajecories of the stochastic interacting particle systems $\disccon$ and $\discann$ (recall Figure \ref{fig:trajs} for one such a sample), we apply the standard Kinetic Monte Carlo approach for continuous-time Markov chain as described in \cite{Voter07}. This approach iterates over the jumps. Given the current time $t$ and the list of dislocation positions $\bL = (L_1,\ldots,L_n)$, we compute the rates $r_{i,\pm}$ from \eqref{ri}, compute the total rate 
\[
  R:=\sum_{i=1}^{n_+}r_{i,+}(\bL) + \sum_{i=1}^{n_-}r_{i,-}(\bL),
\]
sample $\tau$ from the exponential distribution with rate $R$, update time to $t + \tau$, compute the transition probabilities $p_{i,\pm} := r_{i,\pm}(\bL)/R$, sample a specific pair $(j,\chi)$ from these probabilities, and update $\bL$ by moving particle $L_j$ to $L_j + \chi \e$. When simulating \eqref{Pann}, if the site $L_j + \chi \e$ contains a particle $k$ with $b_k = -b_j$, then we remove both particles $j$ and $k$ immediately from the system. This iteration over particle jumps stops when $t$ exceeds $T$. 

This algorithm was implemented in Python using the \texttt{numpy} package \cite{numpy}. To reduce computational cost by a factor $O(n)$, we \textit{update} the values of $r_{i,\pm}(\bL)$ after each jump rather than computing them from scratch. This is possible since particles jump individually, and interactions between particles act pairwise. We note that domain decomposition methods could potentially be used to parallelize the code and yield even greater efficiencies, but these were not pursued here.

\subsection{Finite volume discretisation}
\label{sec:FVdiscretisation}
To discretize the PDE systems \eqref{Pcon} and \eqref{Pann}, we apply time-explicit upwind finite volume schemes. The singularities inherent in the velocity field $v$ are dealt with by a simple approach, which appears to work well in practice; see \cite{ApisornpanichVanMeurs23}. Again, the algorithms described below were implemented using \texttt{numpy}.

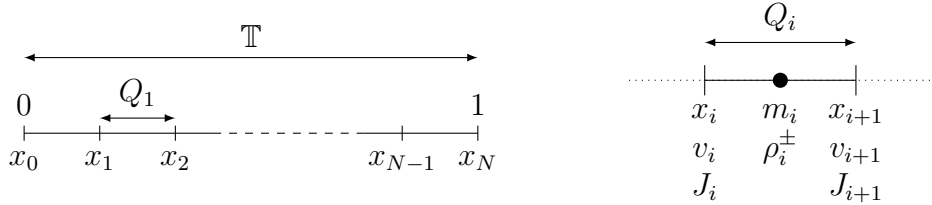
\begin{figure}[h]
 \centering
   \begin{tikzpicture}[scale=1, >= latex]
     \def \a {9} 

     \draw[<->] (0,1) -- node[above] {$\T$} (6,1);
     \draw (0,0)  --++ (2.5,0);
     \draw[dashed] (2.5,0) -- (4.5,0);
     \draw (4.5,0) -- (6,0);
     \draw (0,-.1) node[below]{$x_0$} --++ (0,.2) node[above]{$0$}; 
     \draw (1,-.1) node[below]{$x_1$} --++ (0,.2);
     \draw[<->] (1,.2) --++ (1,0) node[midway, above]{$Q_1$};
     \draw (2,-.1) node[below]{$x_2$} --++ (0,.2) ;
     \draw (5,-.1) node[below]{$x_{N-1}$} --++ (0,.2);
     \draw (6,-.1) node[below]{$x_N$} --++ (0,.2) node[above]{$1$};    
     
     \begin{scope}[shift={(9,.7)}]
       \draw[dotted] (-1,0) -- (3,0);
       \draw (0,0) -- (2,0);
       \draw (0,0) node[below]{\begin{tabular}{c} $x_i$ \\ $v_i$ \\ $J_i$ \end{tabular}};
       \draw[<->] (0,0.5) -- node[above] {$Q_i$} (2,0.5);
       \draw (0,-.2) --++ (0,.4); 
       \fill (1,0) circle (.1) node[below]{\begin{tabular}{c} $m_i$ \\ $\rho_i^\pm$ \end{tabular}};
       \draw (1,-.1); 
       \draw (2,-.2)  --++ (0,.4) node[midway, below]{\begin{tabular}{c} $x_{i+1}$ \\ $v_{i+1}$ \\ $J_{i+1}$ \end{tabular}};
     \end{scope}
   \end{tikzpicture}
   \caption{Left: discretization of $\T$. Right: a cell $Q_i$, including the places where several quantities are defined or interpreted.}
   \label{fig:FVdiscretisation}
 \end{figure}

The discretization of $\T$ is illustrated in Figure~\ref{fig:FVdiscretisation}. Precisely, we interpret $\T$ as $[0,1]$ and discretize it by $N+1$ points $x_i = \frac{i}{N}$ for $i=0,1,\ldots,N$. We consider the intervals $Q_i := (x_i,x_{i+1})$ for $i = 0,1,\ldots,N-1$ as the finite volume cells with midpoints $m_i := \frac12 (x_i + x_{i+1})$.

Let $\rho^\pm$ and $\kappa$ be the exact solutions of $\contcon$ and $\contann$ respectively.
The mean dislocation densities and the mean net Burgers vector on cell $i$ are represented by
 \[
   \rho_i^\pm := \frac1{|Q_i|} \int_{Q_i} \rho^\pm \,\mathrm{d}x=N \int_{Q_i} \rho^\pm \,\mathrm{d}x
   \qquad \text{and} \qquad
   \kappa_i := N \int_{Q_i} \kappa \,\mathrm{d}x
 \]
respectively.

Next we introduce the schemes. We use the scheme from \cite{ApisornpanichVanMeurs23} for \eqref{Pann}. For \eqref{Pcon} we use a modification of it. Since the modification is simpler than the original scheme, we start with the former.

 \paragraph{The scheme for \eqref{Pcon}.} Integrating the equations \eqref{Pcon} over a given cell and using the fundamental theorem of calculus we obtain
 \begin{equation}\label{pfzz}
   \begin{aligned} 
  \partial_t \rho_i^+ 
  &= +N \big[ \rho^+ (x_i) v[\tilde \kappa](x_i) - \rho^+ (x_{i+1}) v[\tilde \kappa](x_{i+1}) \big],\\
  \partial_t \rho_i^- 
  &= -N \big[ \rho^- (x_i) v[\tilde \kappa](x_i) - \rho^- (x_{i+1}) v[\tilde \kappa](x_{i+1}) \big],
\end{aligned}
\end{equation}
where $\tilde \kappa := \rho^+ - \rho^-$ should not be confused with the solution $\kappa$ of $\contann$.
It remains to devise an approximation to the fluxes $J^\pm := \rho^\pm v$ at the ends of the cells. For this we treat $v[\tilde \kappa](x_i)$ and $\rho^\pm (x_i)$ separately. Formally, we compute
\begin{equation*}
\begin{aligned}
  v[\tilde \kappa](x_i) = \int_\T f(x_i-y) \tilde \kappa (y) \, dy
  &= \sum_{j=0}^{N-1} \int_{Q_j} f(x_i-y) \tilde \kappa (y) \, dy \\
  &\approx \frac1N \sum_{j=0}^{N-1} f(x_i - m_j) \tilde \kappa_j =: v_i,
\end{aligned} 
\end{equation*}
where we replace the integrals over the cells by a midpoint approximation. This approximation is designed so that the singularity of $f$ `cancels out'. We refer to \cite[Section 3.4]{ApisornpanichVanMeurs23} for a detailed discussion.

To approximate the values of the densities at the cell endpoints $\rho^+(x_i)$, we apply the upwind scheme: if $v_i \geq 0$, then mass flows to the right, and thus we approximate $\rho^+(x_i) \approx \rho_{i-1}^+$ as the information from the left. If $v_i<0$, we set $\rho^+(x_i) \approx \rho_i^+$. 

Substituting these approximation in \eqref{pfzz}, we obtain
\begin{equation} \label{pfzy}
  \begin{aligned}
  \partial_t \rho_i^+ 
  \approx N \big( J_i^+ - J_{i+1}^+ \big), \qquad 
  J_j^+ &:= \begin{cases}
    +\rho_{j-1}^+ v_j
    &\text{if } v_j \geq 0  \\
    +\rho_j^+ v_j
    &\text{if } v_j < 0, 
  \end{cases}\\
   \partial_t \rho_i^- 
  \approx N \big( J_i^- - J_{i+1}^- \big), \qquad 
  J_j^- &:= \begin{cases}
    -\rho_{j-1}^- v_j
    &\text{if } v_j \leq 0 \\
    -\rho_j^- v_j
    &\text{if } v_j >0,\\ 
  \end{cases}
  \end{aligned}
\end{equation}
where $J_i^+$ represents the approximation of the flux of $\rho^+$ through the cell endpoint $x_i$, and analogously $J_i^-$ represents the approximation of the flux of $\rho^-$ through the same point. 
The scheme for the spatial discretisation is given by \eqref{pfzy} upon replacing the approximation signs above by equalities and by treating $\rho_i^\pm : [0,T] \to [0,\infty)$ as defined through these equations.

To integrate this scheme in time, we use an explicit Euler method with a fixed time step $h = 1/(10 N \log N)$. For this time step the scheme appears to be stable; see \cite[Remark 3.4]{ApisornpanichVanMeurs23}.

\paragraph{The scheme for \eqref{Pann}.} Analogously to \eqref{Pcon}, we  integrate \eqref{Pann} over $Q_i$. This yields
 \begin{equation} \label{pfzv}
  \partial_t \kappa_i 
  = N \big[ |\kappa (x_i)| v[ \kappa](x_i) - |\kappa (x_{i+1})| v[ \kappa](x_{i+1}) \big].
\end{equation}
Fix a time point $t_k$. If $\kappa_i(t_k), \kappa(t_k, x_i), \kappa(t_k, x_{i+1}) > 0$ (note that all three values are expected to be roughly equal), then \eqref{pfzv} implies
\begin{equation*}
  \partial_t [\kappa_i(t_k)]_+
  = N \big( [\kappa (t_k, x_i)]_+ v[\kappa](x_i) - [\kappa (t_k, x_{i+1})]_+ v[ \kappa](x_{i+1}) \big)
\end{equation*}
and $\partial_t [\kappa_i(t_k)]_- = 0$, 
where we recall the positive and negative parts $[\cdots]_\pm$ from \eqref{posneg:part}. Similarly, if $\kappa_i(t_k), \kappa(t_k, x_i),$ $ \kappa(t_k, x_{i+1}) < 0$, then $\partial_t [\kappa_i(t_k)]_+ = 0$ and \eqref{pfzv} implies
\begin{equation*}
  -\partial_t [\kappa_i(t_k)]_-
  = N \big( [\kappa (t_k, x_i)]_- v[\kappa](x_i) - [\kappa (t_k, x_{i+1})]_- v[ \kappa](x_{i+1}) \big).
\end{equation*}
In either of the two cases, we can rewrite the system as
\begin{equation*} 
   \begin{aligned} 
  \partial_t [\kappa_i(t_k)]_+
  &= N \big( [\kappa (t_k, x_i)]_+ v[\kappa](x_i) - [\kappa (t_k, x_{i+1})]_+ v[ \kappa](x_{i+1}) \big),\\
  \partial_t [\kappa_i(t_k)]_-
  &= -N \big( [\kappa (t_k, x_i)]_- v[\kappa](x_i) - [\kappa (t_k, x_{i+1})]_- v[ \kappa](x_{i+1}) \big).
\end{aligned}
\end{equation*}
Note that this is the same set of equations as in \eqref{pfzz}.  

To introduce the scheme, we set $\tilde \rho_i^{k,\pm} := [\kappa_i (t_k)]_\pm$ and $\tilde \rho^{k,\pm} (x_j) := [\kappa (t_k, x_j)]_\pm$. We apply the same time step update as done in the scheme for $\contcon$ to compute $\tilde \rho_i^{k+1,\pm}$. Finally, we set
\begin{equation} \label{pfzt}
  \kappa_i(t_{k+1}) := \tilde \rho_i^{k+1,\pm} - \tilde \rho_i^{k+1,\pm}
\end{equation}
to complete the time step update for \eqref{Pann}. We remark that in the step \eqref{pfzt} of the scheme the annihilation takes place. Indeed, while $\tilde \rho_i^{k,\pm}$ have disjoint support (i.e.\ $\tilde \rho_i^{k,+} \tilde \rho_i^{k,-} = 0$ for all $i$), this need not be true for $\tilde \rho_i^{k+1,\pm}$. If not, then by applying \eqref{pfzt} and afterwards taking the positive and negative part of $\kappa_i(t_{k+1})$, mass has been removed from the system.

\paragraph{Scheme properties.} The simulations in \cite{ApisornpanichVanMeurs23} of \eqref{Pann} on $\R$ suggest that the discretisation error is linear in the cell length $\frac1N$. We have verified numerically that the same accuracy appears to hold for our schemes of \eqref{Pcon} and \eqref{Pann} on $\T$. Moreover, in \cite[Section 3]{ApisornpanichVanMeurs23} it is proven that $\sum_i \kappa_i$ is conserved and that $\sum_i |\kappa_i|$ is decreasing in time. A similar proof shows that this remains true for our setting on $\T$. As our scheme of \eqref{Pcon} follows precisely the finite volume method, $\sum_i \rho^+_i$ and $\sum_i \rho^-_i$ are both conserved.

\subsection{Initial conditions}
\label{sec:method:ICs}
In order to compare the lists $\bL$ computed by the Kinetic Monte Carlo approach and the mean density values $\rho_i^\pm$ and $\kappa_i$ computed from the finite volume schemes, we must make a comparable choice of initial conditions. We do this by choosing periodic density functions $\rho^{\circ,\pm}:\T\to[0,+\infty)$, and deterministically sampling $\bL(0)$, $\rho_i^\pm(0)$ and $\kappa_i(0)$ from it. For the latter two, we take
\[
  \rho_i^\pm(0) = N \int_{Q_i} \rho^{\circ,\pm}, \qquad
  \kappa_i(0) = N\int_{Q_i}\rho^{\circ,+}(x)-\rho^{\circ,-}(x)\,\mathrm{d}x.
\]
We take $\bL(0)$ and $\bb$ as follows. First, we compute the cumulative density functions
\[
  F^{\circ,\pm}(x) = \int_0^x\rho^{\circ,\pm}(y)\,\mathrm{d}y.
\]
Then, for each $j = 1,\ldots,n^\pm$ we solve $F^{\circ,\pm}(y) = \frac 1n(j - \frac12)$, and put a dislocation with the appropriate sign at $\e k \in \Lambda_\e$, where $k \in \N$ is the smallest integer satisfying $m_k \geq y$. With this choice, we have for $N$ sufficiently large that $\int_{L_i}^{L_j} \rho^{\circ,b} dx \approx \frac1n$ for each pair $(i, j)$ of consecutive dislocations of the same sign $b$. 

\subsection{Comparing discrete positions to densities}
\label{sec:method:comparison}
To compare trajectories of dislocation positions $\bL$ with densities $\rho_i^\pm$ or $\kappa_i$ at a given time point, we construct an appropriate metric. 
As a first step, we describe $\bL$, $\rho_i^\pm$ and $\kappa_i$ as (signed) empirical measures. For $\bL$ we introduce the signed discrete density
\begin{equation} \label{nun} 
\nu_n := \frac{1}{n}\left(\sum_{i=1}^{n^+}\delta_{L^+_i}-\sum_{i=1}^{n^-}\delta_{L^-_i}\right),
\end{equation}
where $n^+$ and $n^-$ are the number of positive and negative dislocations respectively. We interpret this choice as each dislocation carrying the fraction $\frac1n$ of the total Burgers vector mass. For $\rho_i^\pm$  and $\kappa_i$ we take
\[
  \mu^\pm := \frac{1}{N}\sum_{j=1}^N \rho_j^\pm \delta_{m_j}, \qquad
  \nu :=\frac{1}{N}\sum_{j=1}^N\kappa_j\delta_{m_j},
\]
where we recall that $N$ is the number of finite volume cells. The signed density corresponding to $(\rho_i^+, \rho_i^-)$ is $\mu^+ - \mu^-$.

In the second step we must choose a metric to quantify the distance between $\nu_n$ and $\nu$ or between $\nu_n$ and $\mu^+ - \mu^-$. We focus on the former, and apply the same metric to the latter. We use the modified version of the $1$-Wasserstein distance introduced in \cite{Mainini12Wdist}, which is given by
\begin{equation} \label{W:Mainini} 
  \W(\nu_n, \nu) := W_1 \big([\nu_n]_+ + [\nu]_-, [\nu_n]_- + [\nu]_+ \big),
\end{equation}
where $W_1$ is the usual $1$-Wasserstein distance, and $[\cdot]_{\pm}$ denotes the positive or negative part of the measure. For non-negative measures $\mu_1, \mu_2$ on $[0,1]$ (instead of $\T$) of equal mass $A \geq 0$, the $1$--Wasserstein distance can be expressed as \cite[Remark 2.19(ii)]{Villani03}
\begin{equation} \label{Wp:on:01}
  W_1(\mu_1, \mu_2) = \int_0^{A} \big| \xi_1(s) - \xi_2(s) \big| ds,
\end{equation}
where $\xi_i : [0,1] \to [0,1]$ are the pseudo-inverses of the cumulative distributions $F_i(x) := \mu_i([0,x])$. Note however that on $\T$ the value of the distance may be lower, as mass can be freely transported across the periodic boundary. It is shown in \cite{RabinDelonGousseau11} that the formula \eqref{Wp:on:01} still applies after an appropriate spatial shift is applied to the measures on $\T$. In practice, this leads to
\begin{equation} \label{W:on:T}
  W_1(\mu_1, \mu_2) := \inf_{z \in \T} \int_0^A \big| \xi_1^z(s) - \xi_2^z(s) \big| ds,
\end{equation}
where $\xi_i^z$ are the pseudo-inverses of the cumulative distribution functions $F^z_i(x) = \mu_i([z,z+x])$.

In our application of \eqref{W:on:T} to \eqref{W:Mainini}, the measures considered are empirical measures with discrete support. As such, $\xi_i$ are step functions, and thus the integral in \eqref{W:on:T} can be computed exactly. Moreover, solving the minimisation over $z$ only requires us to check a finite number of candidate gridpoints; in practice, we find the approximate minimum by finding the minimal value over all $z\in\{x_0,x_1,\ldots,x_{N-1}\}$.

We now make a few remarks about the choice of $\W$ as a metric. First, optimal transport distances require measures of equal total mass. Indeed, to make sense of the definition \eqref{W:Mainini}, the right-hand side requires both (non-negative) input measures to have the same total mass. This holds in our case. Indeed, our choice of the initial condition results in equality of the total net Burgers vector, i.e.\ $\nu_n(\T) = \nu(\T)$, and our schemes preserve this in time. Our reason for taking the $1$-Wasserstein distance as opposed to the common $2$-Wasserstein distance, is that the former is an actual distance whereas the latter is not \cite[Proposition 3.9]{Mainini12Wdist}. 

Second, $\W(\nu_n, \nu)$ allows for `self-annihilation', i.e.\ transport between $[\nu]_+$ and $[\nu]_-$ (same for $\nu_n$). It is not clear whether this is physically meaningful. Yet, since the masses of $[\nu_n]_+$ and $[\nu]_+$ need not be equal (same for the negative parts), one has to take into account the surplus in some manner.

Third, a completely different metric would be to consider $\| \nu_n - \nu \|_{H^{-s}(\T)}$ for $s \geq \frac12$. Since $\T$ is periodic and the measures are empirical, this can easily be computed numerically by using the Fourier transform. However, the physical interpretation of this norm with respect to our models is unclear to us.

Fourth, we have implemented all aforementioned norms in Section \ref{sec:results} on the simulations. Qualitatively, there was no significant difference.

\subsection{Trajectory sampling}
\label{sec:method:sampling}

Since trajectories of our discrete models $\bL$ are random, the distance value $W := \W(\nu_n, \nu)$ is a random variable. To evaluate convergence, we must therefore run independent samples of the full dynamics for each set of parameters we consider, providing $M$ samples $\nu_{n,i}$ and corresponding $W_i=\W(\nu_{n,i}, \nu)$ for $i = 1,\ldots,M$. A simple choice would be to directly sample an estimator of $\E[W]$ numerically. However, an inspection of the simulated samples $W_i$ indicates that they are well-modelled as lognormal random variables with parameters $\mu$ and $\sigma$, i.e. $\log W_i \sim \mathcal{N}(\mu,\sigma^2)$. This is natural for random data that represents a (small) distance which must always be positive.

For each set of parameters, we compute estimators for the corresponding $\mu$ and $\sigma$ from the iid random variables $\{W_i\}_{i=1}^M$ as follows:
\begin{equation} \label{m:sigma}
  \hat{\mu} := \frac1M \sum_{i=1}^M \log W_i, \qquad
  \hat{\sigma} := \sqrt{ \frac1{M-1} \sum_{i=1}^M \big| \log W_i - \hat{\mu} \big|^2 }.
\end{equation}
These are the standard statistical formulae for iid samples of the random variable $\log W$. Finally, our value of interest is
\begin{equation} \label{w}
  w := e^{\hat{\mu}},
\end{equation}
which is the statistical estimate of the \textit{median} value of $W = \W(\nu_n, \nu)$.

\subsection{Quantifying discrete-to-continuum convergence}
\label{sec:method:randomness}

Here we explain our methodology for quantifying discrete-to-continuum convergence, which we formally denote as `$\disccon \to \contcon$' and `$\disccon \to \contcon$' as $n \to \infty$. We will test this only for a single initial condition and at a single time $t$, which we take as the end time $T$.
For both the conservation and annihilation models, we interpret discrete-to-continuum convergence as $\E [\W(\nu_n, \nu)] \to 0$ as $n \to \infty$. Note that this is a strong notion of convergence of the trajectories; it implies that almost every sample of the discrete solution is close to the deterministic continuum solution. Therefore, our concept of convergence is of stochastic-to-deterministic type. This is different from the statistical mechanics approach, where the averaged particle distribution obtained over many different samples of discrete solutions is used to test convergence to the continuum solution.

Numerically, we test our hypothesis that $\E [\W(\nu_n, \nu)] \to 0$ by plotting $w$ (recall \eqref{w}) versus $n$ and checking that $\sigma$ remains bounded. If $w$ tends to $0$, then we interpret this as numerical evidence for convergence. Otherwise, we interpret it as numerical evidence for failure of convergence to the specified limit. We denote this by `$\disccon \not \to \contcon$' or `$\disccon \not \to \contcon$'.

To quantify whether the numerical values of $w$ are large or small, we compare them to the reference value \begin{equation} \label{w:infty}
  w_\infty := \W(\rho^+ - \rho^-, \kappa),
\end{equation} 
which is the distance between the solutions of \eqref{Pcon} and \eqref{Pann}. 
Thus, if $w$ is (well) below $w_\infty$, then we can conclude that the discrete solution is (much) closer to the corresponding continuous model than to the other continuum model, but otherwise no conclusion can be drawn as to which of the models $\contcon$ or $\contann$ fits best.

\section{Computational results}
\label{sec:results}

Computational results were obtained using the cluster resources provided through the University of Warwick's Scientific Computing Research Technology Platform. In particular, as described in Section \ref{sec:method}, ensembles of trajectories for the discrete models were simulated over a fixed time period under a range of parametric assumptions. The particle distributions were then compared with a finite-volume discretisation of the corresponding mean-field PDE models.

\subsection{Choice of the input data for $\contcon$ and $\contann$}
\label{sec:results:PDEcomparison}
\begin{figure}[h!]
  \centering
  \begin{tikzpicture}    
      \draw (0,0) node{\includegraphics[width = .4\textwidth]{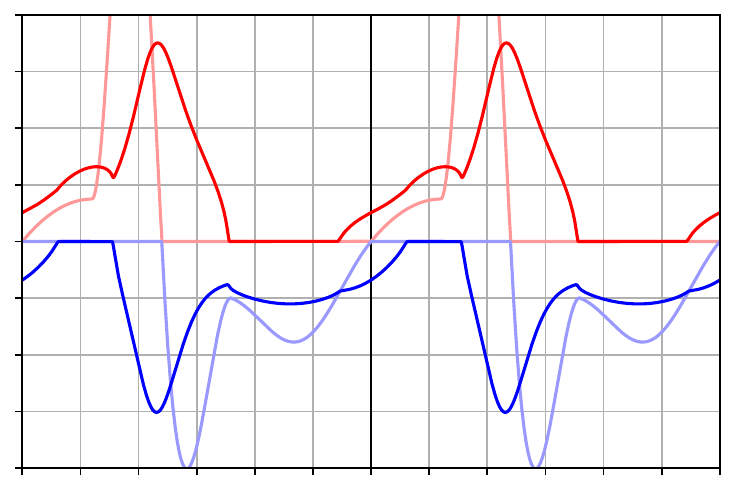}};
      \draw (-3.05,-2) node[below]{$0$};
      \draw (-1.5,-2) node[below]{$\frac12$};
      \draw (.05,-2) node[below]{$1$};
      \draw (1.6,-2) node[below]{$\frac32$};
      \draw (3.1,-2) node[below]{$2$};
      \draw (-3.1,-2) node[left]{$-2$};
      \draw (-3.1,-1) node[left]{$-1$};
      \draw (-3.1,0) node[left]{$0$};
      \draw (-3.1,1) node[left]{$1$};
      \draw (-3.1,2) node[left]{$2$};
      \draw[red] (-1,1) node {$\rho^+$};
      \draw[blue] (-2.5,-1) node[below] {$\rho^-$};
 
 \begin{scope}[shift={(8,0)}]
    \draw (0,0) node{\includegraphics[width = .4\textwidth]{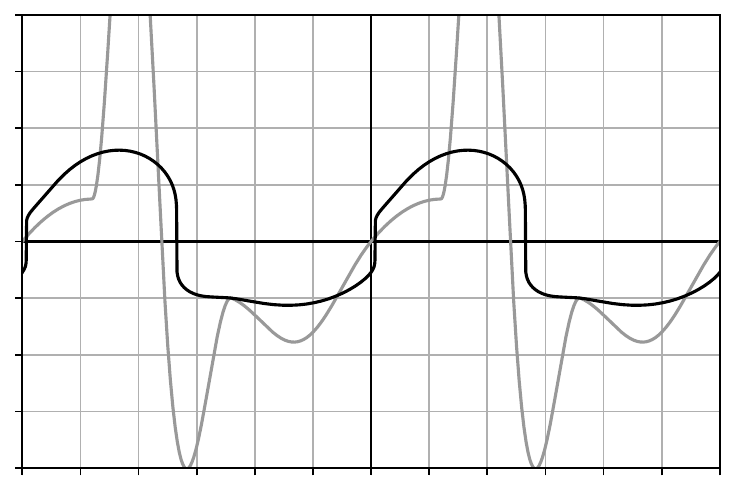}};
      \draw (-3.05,-2) node[below]{$0$};
      \draw (-1.5,-2) node[below]{$\frac12$};
      \draw (.05,-2) node[below]{$1$};
      \draw (1.6,-2) node[below]{$\frac32$};
      \draw (3.1,-2) node[below]{$2$};
      \draw (-3.1,-2) node[left]{$-2$};
      \draw (-3.1,-1) node[left]{$-1$};
      \draw (-3.1,0) node[left]{$0$};
      \draw (-3.1,1) node[left]{$1$};
      \draw (-3.1,2) node[left]{$2$};
   \draw (-1.7,.3) node[above right] {$\kappa$};
 \end{scope}
\end{tikzpicture}
\caption{The solutions $\rho^\pm$ of \eqref{Pcon} and $\kappa$ of \eqref{Pann} at final time, $t = T$. 
  Two periods of $\T$ are shown. The lighter plots represent the initial conditions.
  }
     \label{fig:conSolsSim}
\end{figure}
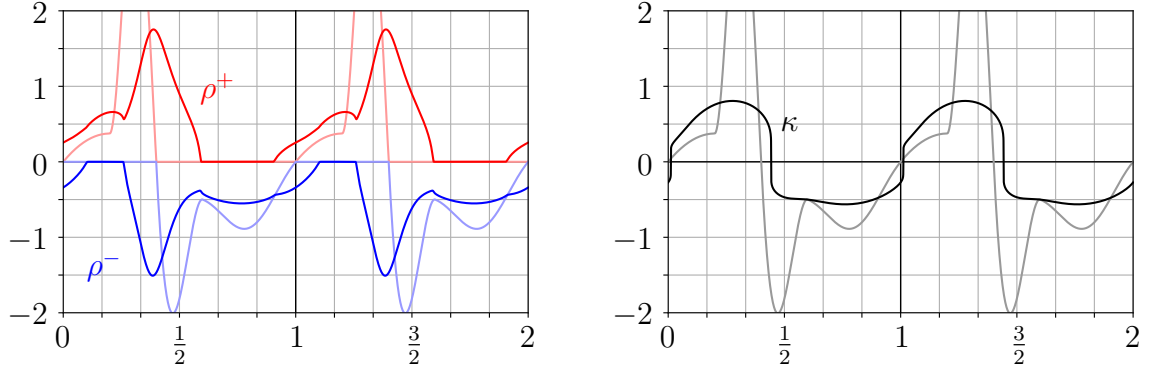

We take the initial condition as follows:
\begin{equation*} 
\begin{aligned}
  \rho^{\circ,+}(x) &:= \tfrac1{10} \rho_1 \left(x; \tfrac25\right) + \tfrac15 \rho_2\left(x - \tfrac15; \tfrac15\right) + \tfrac15 \rho_3\left(x - \tfrac15; \tfrac15\right), \\
  \rho^{\circ,-}(x) &:= \tfrac9{40} \rho_1\left(x - \tfrac25; \tfrac35\right) + \tfrac15 \rho_2\left(\tfrac35 - x; \tfrac15\right) + \tfrac3{40} \rho_3\left(x - \tfrac35; \tfrac25\right),
\end{aligned}
\end{equation*}
where $\rho_i$ are unit mass densities on $\T$ given by
\begin{align*}
  \rho_1(x;b) &:= \frac6{b^3} x(b-x) 1_{0 \leq x \leq b}, \\
  \rho_2(x;b) &:= \frac{12}{b^4} x^2(b - x) 1_{0 \leq x \leq b}, \\
  \rho_3(x; b) &:= \frac2b \sin^2 \left(\frac{\pi x}{b}\right) 1_{0 \leq x \leq b}.
\end{align*}
They are illustrated in Figure \ref{fig:conSolsSim}. 
Our reasons for this choice are:
\begin{itemize}\itemsep0em
\item $\kappa^\circ = \rho^{\circ,+}-\rho^{\circ,-} \in C^1(\T)$, which avoids any spatial regularity issues which otherwise might occur at initial time,
\item $\rho_1$ ensures that $\kappa^\circ$ has full support at initial time;
\item $\rho_2$ leads to a high density close to the front of $\kappa^\circ$ at $x = 0.4$; and
\item $\rho_3$ provides some further asymmetry and ensures that $\kappa^\circ$ has total mass $0$. 
\end{itemize}
We have taken a final time $T = 0.04$ so that $\rho^\pm$ are significantly different from $\rho^\circ$, some absolute Burgers vector density has been annihilated, and $\rho^\pm$ are nowhere near equilibrium, which we expect to be the constant densities. Regarding the numerical parameters, we have taken $N=5120$ cells  and timestep $\tau = 1 / (10 N \log N)$.
 
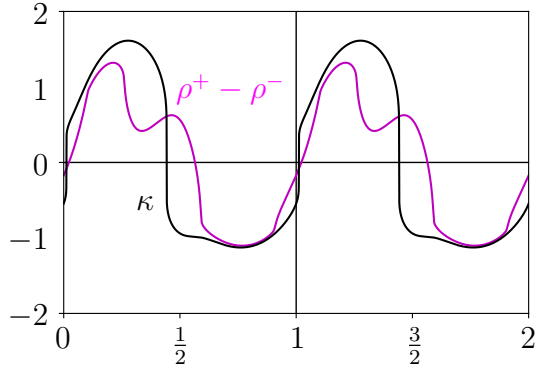
\begin{figure}[h] 
  \centering
   \begin{tikzpicture}
    \def\r {.15}  
    \definecolor{mag}{RGB}{255,0,255}     
      
    \draw (0,0) node{\includegraphics[width = .4\textwidth]{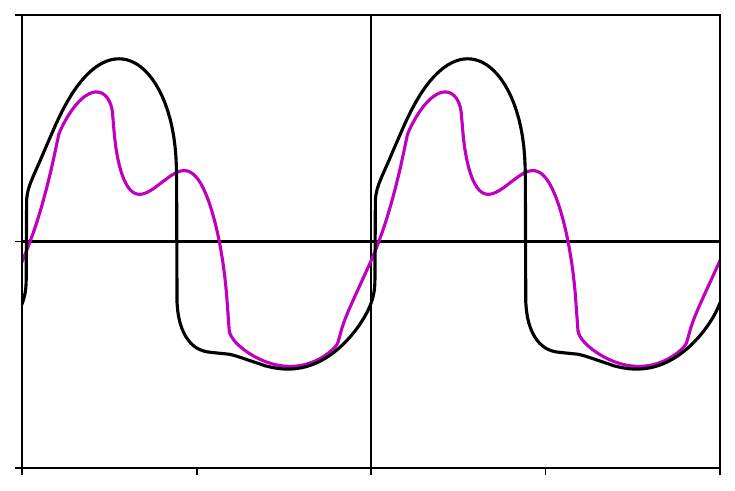}};
    \draw (-3.05,-2) node[below]{$0$};
      \draw (-1.5,-2) node[below]{$\frac12$};
      \draw (.05,-2) node[below]{$1$};
      \draw (1.6,-2) node[below]{$\frac32$};
      \draw (3.1,-2) node[below]{$2$};
      \draw (-3.1,-2) node[left]{$-2$};
      \draw (-3.1,-1) node[left]{$-1$};
      \draw (-3.1,0) node[left]{$0$};
      \draw (-3.1,1) node[left]{$1$};
      \draw (-3.1,2) node[left]{$2$};
    \draw (-1.7,-.5) node[left] {$\kappa$};
    \draw[color=mag] (-1.7,1) node[right] {$\rho^+ - \rho^-$};
 \end{tikzpicture} 
    \caption{The signed densities $\kappa$ and $\rho^+ - \rho^-$ at final time $T$ from Figure \ref{fig:conSolsSim}.}
    \label{fig:kap2}
  \end{figure}

Figure~\ref{fig:conSolsSim} illustrates the solutions, and Figure~\ref{fig:kap2} provides a comparison of the signed densities.
Due to the asymmetry in the initial condition, we see two more features than those discussed when considering Figure~\ref{fig:rho:kap:intro}. First, the point where $\kappa$ changes sign has moved into the support of $[\kappa^\circ]_-$. This makes sense in light of the observation that $\rho^{\circ,+}$ has more mass close to the left of the front than $\rho^{\circ,-}$ has to the right. Second, $\kappa$ again has a steep interface at the point of the sign change, even though $\kappa^\circ$ is regular at the same point. Inspecting Figure~\ref{fig:kap2} shows that the differences between the signed density evolutions are most apparent around the points where $\kappa$ changes sign, reflecting our hypothesis that $\contcon$ and $\contann$ treat annihiliations differently.

  \subsection{Asymptotic choices of parameters}
\label{sec:results:exponents}
 
\begin{figure}[t!]
  \centering
  \begin{tikzpicture}[scale = 1.0]
   \def\r {.04}      
      
   \fill[black!10!white] (0,0) --++ (2,0) --++ (0,2) --++ (1.9,1.9)  --++ (-3.9,0) -- cycle;   
   \draw[black!30!white] (0,0) grid (3.9, 3.9);
    
   \draw[->] (0,-0.1) -- (0.0,4) node[left]{$\alpha_\beta$};
   \draw[->] (-0.1,0) -- (4,0) node[below]{$\alpha_\e$};
   \draw (0,0) -- (3.9,3.9);
   \draw (2,0) -- (2,3.9);
    
   \draw (2,0) node[below]{$1$};
   \draw (3,0) node[below]{$1.5$};
   \draw (0,1.6) node[left]{$0.8$} --++ (.1,0);
   \draw (0,2) node[left]{$1$};
   \draw (0,3) node[left]{$1.5$};
   
   \fill (2,1.6) circle (\r);
   \fill (2.2,1.6) circle (\r);
   \fill (2.4,1.6) circle (\r);
   \fill[blue] (3,1.6) circle (1.5*\r);
   \fill (3,2) circle (\r);
   \fill (3,1) circle (\r);
   \fill (3,2.6) circle (\r);
   \fill (3,2.8) circle (\r);
   \fill (3,3) circle (\r);
\end{tikzpicture}
  \caption{An illustration of our choices (the dots) for the exponents $\alpha_\e$ and $\alpha_\beta$ in $\e \approx n^{-\alpha_\e}$ and $\beta = n^{\alpha_\beta}$. In the grey regions we do not expect convergence of our discrete models to our continuous models (see Section \ref{sec:models:parameters}).}
     \label{fig:epsbeta}
\end{figure}
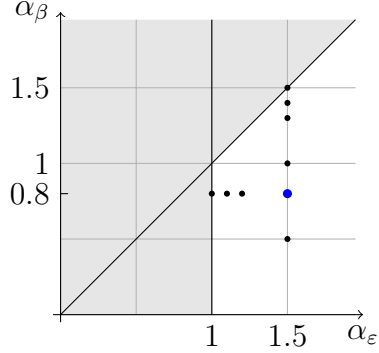

Recall that we choose to investigate the asymptotic parameter regime for $(\e,n,\beta)$ in which $1\ll n, \beta \ll\frac1\e$, as motivated in Section \ref{sec:models:parameters}. In our computations, we have to restrict ourselves to a limited number of points in this parameter space. We choose these by varying $n$, and making the choices
\begin{equation*} 
  \e := \frac1{[n^{\alpha_\e}]}, \qquad \beta := n^{\alpha_\beta},
\end{equation*}
where $[\,\cdot\,]$ indicates rounding to the closest integer, and $\alpha_\e, \alpha_\beta > 0$ are exponents for which we consider various values. The parameter restrictions in Section \ref{sec:models:parameters} translate to $0 < \alpha_\beta < \alpha_\e < 1$. The corresponding regime is plotted in Figure \ref{fig:epsbeta}.

The specific choices of $(\alpha_\e, \alpha_\beta)$ are indicated by the dots in Figure \ref{fig:epsbeta}. These points were chosen  as it was desirable to investigate several points where $\alpha_\e$ is fixed, so that the effect of $\alpha_\beta$ can be observed, and similarly to investigate several points where $\alpha_\beta$ is fixed. These exponents are further chosen to approach the boundaries of the region where our hypotheses on the asymptotic ordering of parameters discussed in Section~\ref{sec:models:parameters} break down, leading to an assessment of the robustness of any convergence behaviour observed across the region.

The number of points in the parameter space were limited to avoid excessive runtime and computational cost: indeed, under the assumption that during the evolution $F_i(\bL)$ is bounded uniformly in $\e, n$, the computational cost for solving $\disccon$ and $\discann$ is
\begin{equation} \label{comp:time}
  O \Big(\frac{n^2}{\beta \e^2} \Big)
  = O (n^{2 + 2 \alpha_\e - \alpha_\beta } )
  \quad \text{as } n \to \infty.
\end{equation} 
To see this, we focus on $\disccon$. The expected runtime is the product of the following three factors:
\begin{enumerate}
  \item $(\beta \e^2)^{-1}$, the expected order of magnitude of a typical jump rate $r_{\pm, i}(\bL)$, which yields the expected number of jumps per unit of simulated time;
  \item $2n$, the number of possible jumps from any state $\bL$; and
  \item $n$, which is the approximate computational cost for updating the rates.
\end{enumerate}
Based on \eqref{comp:time}, we see that in our parameter regime (see Figure~\ref{fig:epsbeta}) the expected runtime grows rapidly in $n$. Precisely, it is at least $O(n^3)$ (only reached at $\alpha_\e = \alpha_\beta = 1$, which is on the boundary of our open regime).  Hence, our choices of $(\alpha_\e, \alpha_\beta)$ remain relatively close to the point $(1,1)$. Still, the runtime ranges from $O(n^3)$ to $O(n^{4.5})$ across our data points; this rapid scaling in $n$ highlights the need and power of effective PDE models, which is one of the motivations of our study.

\subsection{Numerical results}
\label{sec:results:results} 
In this section, we present the results of our numerical simulations. We let $n$ range from $20$ to $200$. 
For each choice of $(n, \alpha_\e, \alpha_\beta)$ we compute $M=50$ sample trajectories of the solutions to $\disccon$ and $\discann$ and evaluate them at time $T = 0.04$. Then, we compute the estimators $w$ and $\hat \sigma$, reflecting the distribution of the distances between the distributions; recall the definitions \eqref{m:sigma} and \eqref{w}. Formally-speaking, $w$ reflects an estimate of the distance between the discrete and continuum solutions, and $w \mathrm{e}^{\pm2\hat \sigma}$ are the quantiles which cover approximately 95\% of the error distribution. Whenever convenient, we write $\wcon$ or $\wann$ to indicate the model where $w$ is computed from.

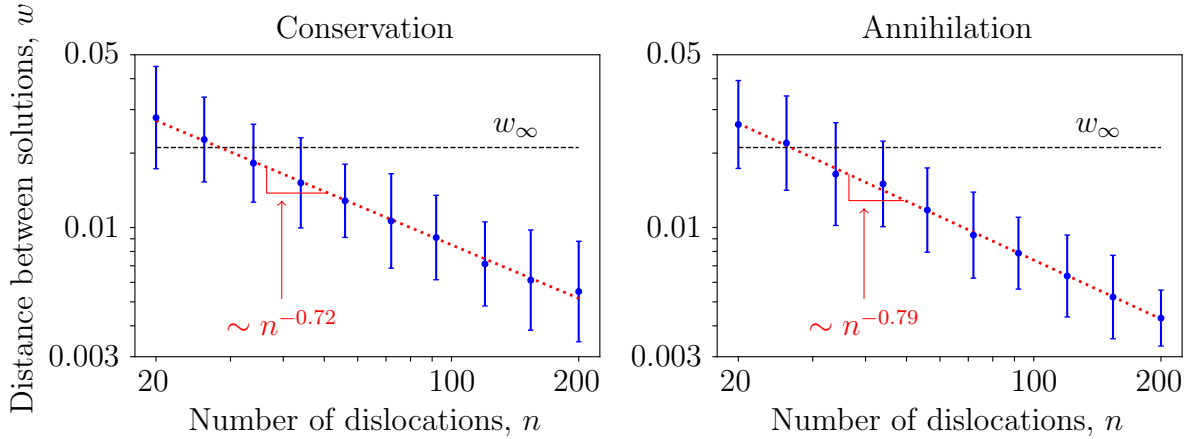
\begin{figure}[h!]
  \centering
  \begin{tikzpicture}
   \def\r {.15}      
      
   \draw (0,-.3) node{\includegraphics[width = .4\textwidth]{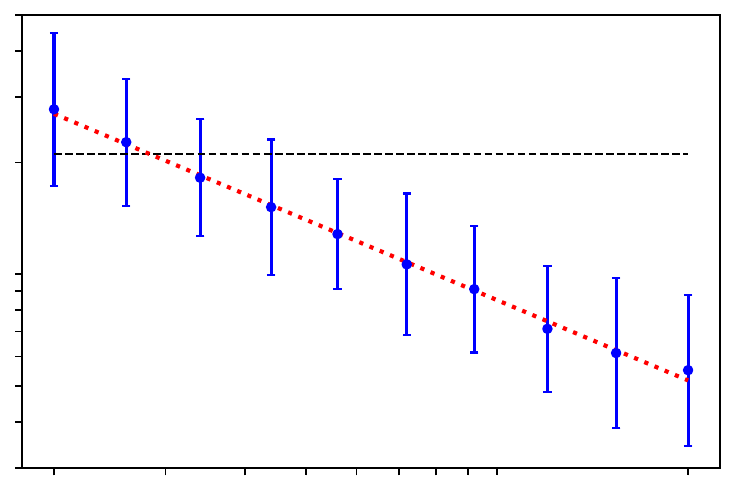}}; 
   \draw[red] (-1.3, .25) -- (-1.3, -.1) -- (-.5, -.1); 
   \draw[red, ->] (-1.1, -1.5) node[below] {$\sim n^{-0.72}$} -- (-1.1, -.2);
   \draw (0,-2.8) node[below]{Number of dislocations, $n$};
   \draw (1.15,-2.3) node[below]{$100$};
   \draw (2.8,-2.3) node[below]{$200$};
   \draw (-2.8,-2.3) node[below]{$20$};
   \draw (0,1.8) node[above]{Conservation};
   \draw (-4.5,-.3) node[rotate=90]{Distance between solutions, $w$};
   \draw (2, .5) node[above] {$w_\infty$};
   \draw (-3.1,-2.25) node[left]{$0.003$};
   \draw (-3.1,-0.55) node[left]{$0.01$};
   \draw (-3.1, 1.75) node[left]{$0.05$};

   \begin{scope}[shift={(7.7, 0)}]  
     \draw (0,-.3) node{\includegraphics[width = .4\textwidth]{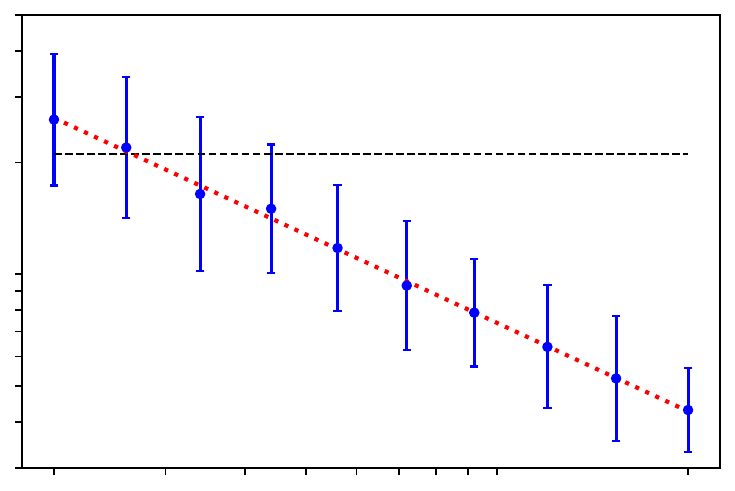}};
   \draw[red] (-1.3, .15) -- (-1.3, -.2) -- (-.57, -.2); 
   \draw[red, ->] (-1.1, -1.5) node[below] {$\sim n^{-0.79}$} -- (-1.1, -.3);
   \draw (1.15,-2.3) node[below]{$100$};
   \draw (2.8,-2.3) node[below]{$200$};
   \draw (-2.8,-2.3) node[below]{$20$};
   \draw (0,-2.8) node[below]{Number of dislocations, $n$};
   \draw (0,1.8) node[above]{Annihilation};
   \draw (2, .5) node[above] {$w_\infty$};
   \draw (-3.1,-2.25) node[left]{$0.003$};
   \draw (-3.1,-0.55) node[left]{$0.01$};
   \draw (-3.1, 1.75) node[left]{$0.05$};
   \end{scope}
\end{tikzpicture}
\caption{Plots of $w$ as a function of $n$ at the reference values $(\alpha_\e, \alpha_\beta) = (1.5, 0.8)$. The blue dots are the values of $w$ and the vertical blue lines are the error bars. The red dotted line is a least-squares power law fit. $w_\infty$ (see \eqref{w:infty}) is added for reference.}
     \label{fig:W:ref:val}
\end{figure}

\paragraph{Results at the reference values.} 
The results of our first convergence investigation are shown in
Figure~\ref{fig:W:ref:val}. We use the reference point $(\alpha_\e, \alpha_\beta) = (1.5, 0.8)$, which correspond to the large dot in Figure \ref{fig:epsbeta}. The plots show that $w$ decays exponentially as $n$ increases, and $\hat{\sigma}$ is constant, so that the quantiles shrink as $n$ increases (note the log-scale on the vertical axis).
Furthermore, within the limited range of $n$, we observe that $w$ and its error bars shrink well below the reference value $w_\infty$ (recall \eqref{w:infty}), which provides strong evidence that the discrete solutions are much closer to their corresponding continuum prediction then to each other. In conclusion, our results suggest that $\disccon \to \contcon$ and $\discann \to \contann$.

In most of the simulations in this section it was found that $\hat \sigma$ does not significantly depend on $n$ and is of similar magnitude across the parameter choices, showing that the distributions cluster tightly around the median as $n$ increases. As such, we only display error bars when significant variation of $\hat \sigma$ occurs. 

\begin{figure}[h!]
  \centering
  \begin{tikzpicture}[scale = .9]
   \def\r {.15}        
      
    \begin{scope}[shift={(0,0)}, scale = .75]
       \draw (0,0) node{\includegraphics[width = .27\textwidth]{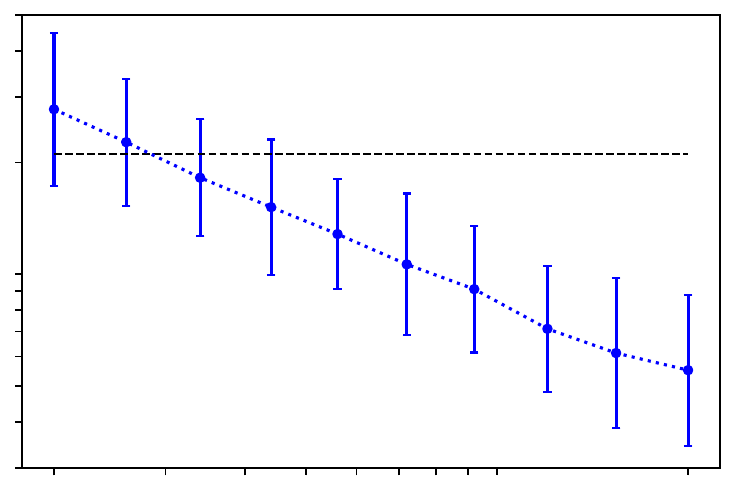}};
       \draw (0,2) node[below] {$\alpha_\beta=0.8$};
       \draw (1.15,-2) node[below]{$100$};
	   \draw (2.8,-2) node[below]{$200$};
	   \draw (-2.8,-2) node[below]{$20$};
	   \draw (-3.1,-1.95) node[left]{$0.003$};
   	   \draw (-3.1,-0.25) node[left]{$0.01$};
   	   \draw (-3.1, 2.05) node[left]{$0.05$};
    \end{scope}
   
   \begin{scope}[shift={(5.3,0)}, scale = .75]
       \draw (0,0) node{\includegraphics[width = .27\textwidth]{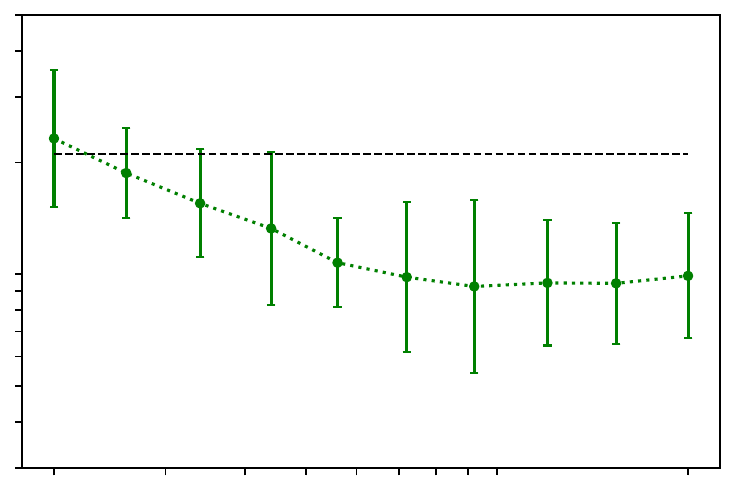}};
       \draw (0,2.1) node[above]{Conservation};
       \draw (0,2) node[below] {$\alpha_\beta=1$};
       \draw (1.15,-2) node[below]{$100$};
	   \draw (2.8,-2) node[below]{$200$}; 
	   \draw (-2.8,-2) node[below]{$20$};
    \end{scope}
    
    \begin{scope}[shift={(10.6,0)}, scale = .75]
       \draw (0,0) node{\includegraphics[width = .27\textwidth]{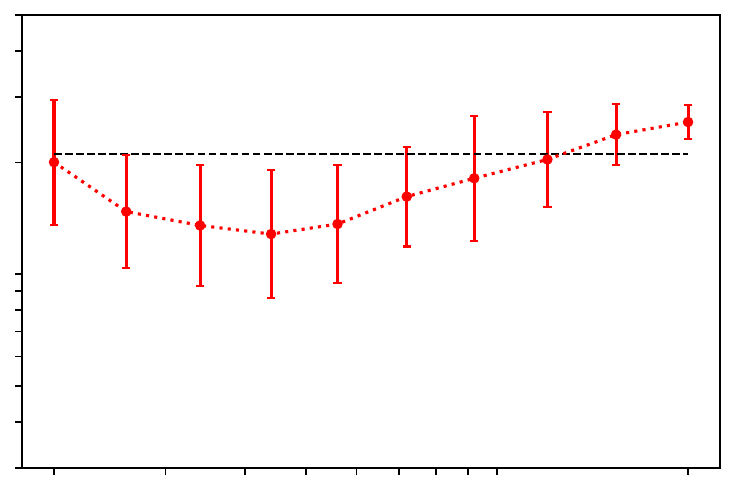}};
       \draw (0,2) node[below] {$\alpha_\beta=1.3$};
       \draw (1.15,-2) node[below]{$100$};
	   \draw (2.8,-2) node[below]{$200$};
	   \draw (-2.8,-2) node[below]{$20$};
    \end{scope}
\end{tikzpicture}
  \caption{Median values of $\wcon$ for $\alpha_\e = 1.5$ with varying $\alpha_\beta$; error bars indicate a 95\% confidence interval. The exponent $\alpha_\beta$ increases from left to right, and we observe a trend of increasing error between the discrete and continuum models for large $n$.}
     \label{fig:W:no-ann}
\end{figure}
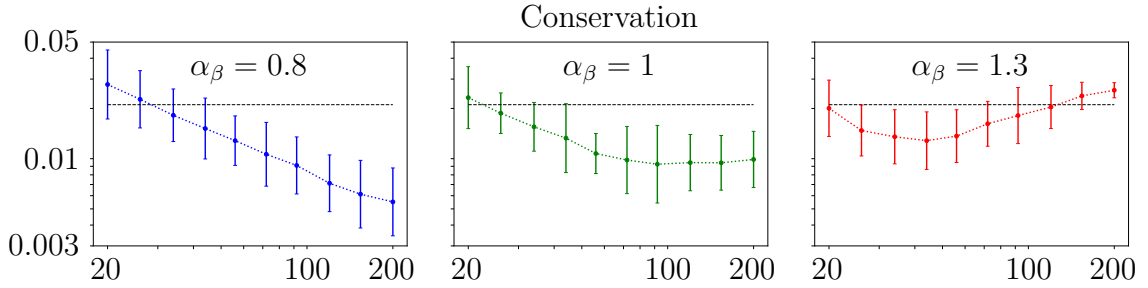

\paragraph{Results for the conservation models.}
We first investigate convergence in the case of the conservative models. Taking $\alpha_\e = 1.5$ to be fixed, we vary the exponent $\alpha_\beta$, with the results shown in
Figure~\ref{fig:W:no-ann}. Clearly, there is a dramatic shift in the behavior of $w$ as $\alpha_\beta$ increases. For the largest value of $\alpha_\beta$, we see that $w$ actually increases (after $n=50$), rising beyond the reference value $w_\infty$. This is not due to random fluctuations. This strongly suggests $\disccon \not \to \contcon$ at $(\alpha_\e, \alpha_\beta) = (1.5, 1.3)$. For $\alpha_\beta=1$ the results are more subtle; the halted decrease of $w$ suggests no convergence, and thus it seems that $\disccon \not \to \contcon$. Still, $w$ and its error bars remain below $w_\infty$, which means that the discrete solutions is  relatively close to the continuous solution of $\disccon$. Finally, when viewing the trend from right to left (i.e.\ $\alpha_\beta$ decreasing), one may even expect that $\disccon \not \to \contcon$ for $(\alpha_\e, \alpha_\beta) = (1.5, 0.8)$, which is \textit{opposite} to what we expected from Figure \ref{fig:W:ref:val}. Therefore, it appears our numerical results are inconclusive as to whether $\disccon$ converges to $\contcon$ or not. To get a decisive answer one would need to go well beyond $n=200$, which is computationally out of reach. Therefore, we did not compute the solution of $\disccon$ for the remaining data points of $(\alpha_\e, \alpha_\beta)$.

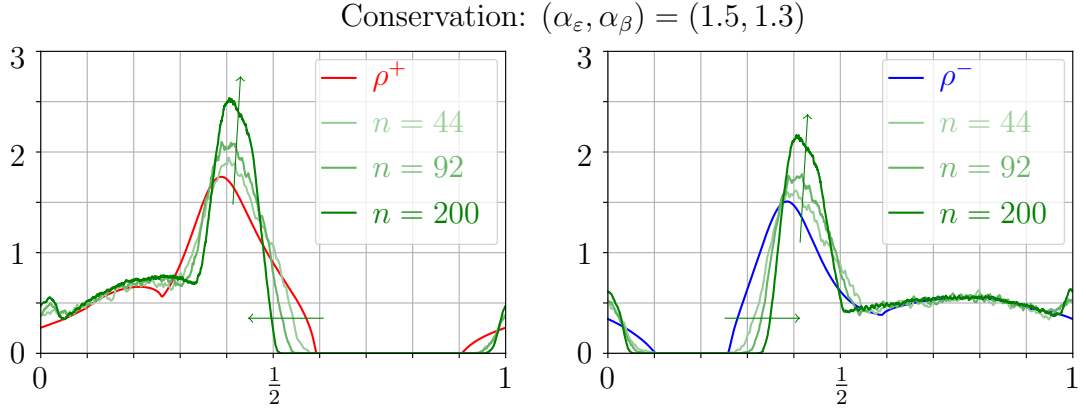
\begin{figure}[h!]
  \centering
  \begin{tikzpicture}
   \def\r {.15}   
   \definecolor{g1}{RGB}{0,127.5,0}  
   \definecolor{g2}{RGB}{102,178.5,102}
   \definecolor{g3}{RGB}{153,205,153} 
   
   \draw (4,2.1) node[above]{Conservation: $(\alpha_\e, \alpha_\beta) = (1.5, 1.3)$};   
   \draw (0,0) node{\includegraphics[width = .4\textwidth]{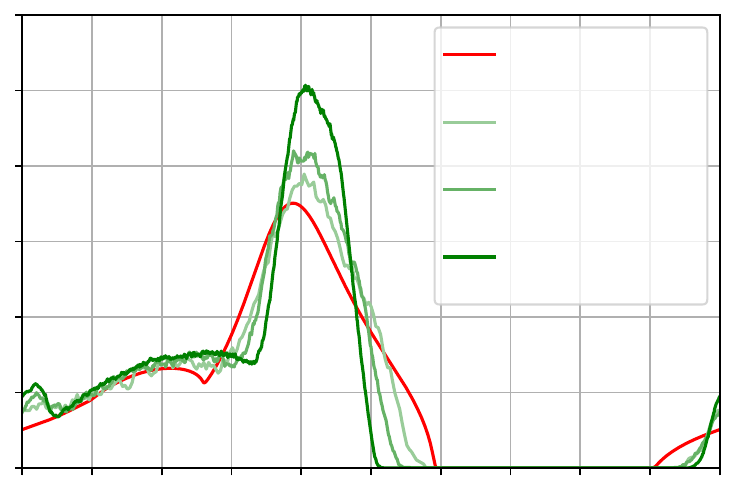}}; 
   \draw (-3.05,-2) node[below]{$0$};
   \draw (.05,-2) node[below]{$\frac12$};
   \draw (3.1,-2) node[below]{$1$};
   \draw (-3.1,-1.95) node[left]{$0$};
   \draw (-3.1,-0.65) node[left]{$1$};
   \draw (-3.1, 0.65) node[left]{$2$};
   \draw (-3.1, 1.95) node[left]{$3$};
   \draw[red] (1.2, 1.7) node[right] {$\rho^+$};
   \draw[g3] (1.2, 1.1) node[right] {$n = 44$};
   \draw[g2] (1.2,  .5) node[right] {$n = 92$};
   \draw[g1] (1.2, -.1) node[right] {$n = 200$};  
   \draw[g1,->] (.7,-1.5) --++ (-1,0);
   \draw[g1,->] (-.5,0) -- (-.4,1.7);
   
   \begin{scope}[shift={(7.5, 0)}]
     \draw (0,0) node{\includegraphics[width = .4\textwidth]{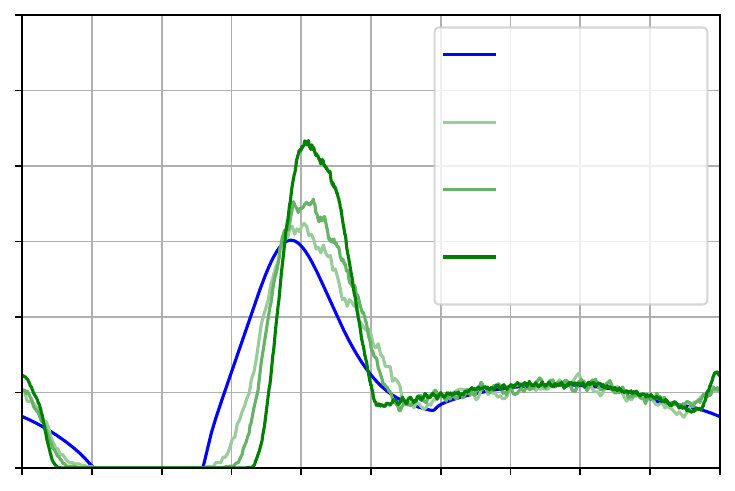}}; 
   \draw (-3.05,-2) node[below]{$0$};
   \draw (.05,-2) node[below]{$\frac12$};
   \draw (3.1,-2) node[below]{$1$};
   \draw (-3.1,-1.95) node[left]{$0$};
   \draw (-3.1,-0.65) node[left]{$1$};
   \draw (-3.1, 0.65) node[left]{$2$};
   \draw (-3.1, 1.95) node[left]{$3$};
   \draw[blue] (1.2, 1.7) node[right] {$\rho^-$};
   \draw[g3] (1.2, 1.1) node[right] {$n = 44$};
   \draw[g2] (1.2,  .5) node[right] {$n = 92$};
   \draw[g1] (1.2, -.1) node[right] {$n = 200$};
   \draw[g1,->] (-1.5,-1.5) --++ (1,0);
   \draw[g1,->] (-.5,-.5) -- (-.4,1.2);
   \end{scope}
\end{tikzpicture}
  \caption{The statistical densities $\lambda_n^+$ (left) and $\lambda_n^-$ (right). $\lambda_n^\pm$ are computed by summing all $M = 50$ samples of $[\nu_n]_+$ and all of $[\nu_n]_-$, mollifying by taking the spatial average over the length $\frac1{20}$, and then normalizing to unit mass.}
     \label{fig:rhopm:stat:no-ann}
\end{figure}

The increasing values of $w$ at $(\alpha_\e, \alpha_\beta) = (1.5, 1.3)$ were unexpected. To investigate what happens to individual discrete solutions, in Figure~\ref{fig:rhopm:stat:no-ann} we plot the \textit{statistical densities} $\lambda_n^+$ and $\lambda_n^-$ of the discrete solutions, which is obtained by summing all samples of $[\nu_n]_+$ and all of $[\nu_n]_-$ (recall \eqref{nun}), renormalizing and mollifying. We observe that away from the fronts around $x = 0$ and $x = 0.4$, $\lambda_n^\pm$ are close to $\rho^\pm$, the solutions of $\contcon$. However, around the fronts, $\lambda_n^\pm$ tend to accumulate more than $\rho^\pm$. Moreover, this accumulation effect gets stronger as $n$ increases, which is in line with an increasing $w$. 

\begin{figure}[h!]
  \centering
  \begin{tikzpicture}
   \def\r {.15}      
   \definecolor{mag}{RGB}{255,0,255}
   \definecolor{g1}{RGB}{0,127.5,0}  
   \definecolor{g2}{RGB}{102,178.5,102}
   \definecolor{g3}{RGB}{153,205,153}
   
   \begin{scope}[scale = 1]   
   \draw (0,2.1) node[above]{Conservation: $(\alpha_\e, \alpha_\beta) = (1.5, 1.3)$};
   \draw (0,0) node{\includegraphics[width = .4\textwidth]{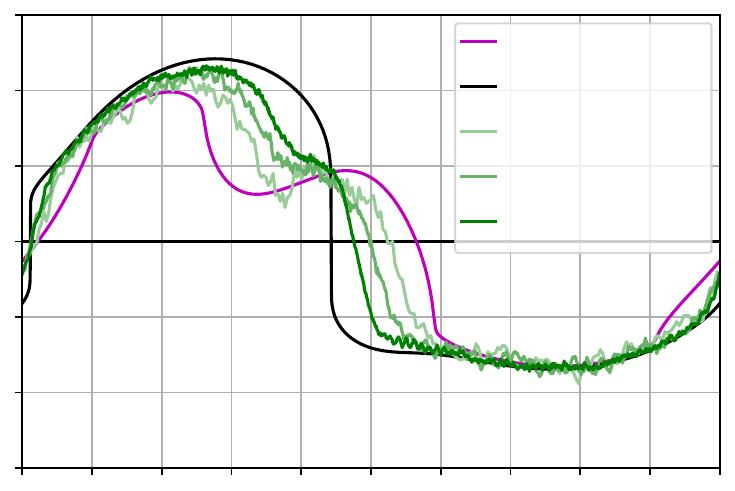}}; 
   \draw (-3.05,-2) node[below]{$0$};
   \draw (.05,-2) node[below]{$\frac12$};
   \draw (3.1,-2) node[below]{$1$};
   \draw (-3.1,-1.95) node[left]{$-1$};
   \draw (-3.1,0) node[left]{$0$};
   \draw (-3.1, 2.0) node[left]{$1$};
   \draw[color=mag] (1.2, 1.8) node[right] { $\rho^+ - \rho^-$};
   \draw (1.2, 1.4) node[right] { $\kappa$};
   \draw[g3] (1.2, 1.025) node[right] { $n = 44$};
   \draw[g2] (1.2,  .625) node[right] { $n = 92$};
   \draw[g1] (1.2, .225) node[right] {$n = 200$};
   \draw[g1,->] (-1.2,0.8) -- (-.7,1.3);
   \draw[g1,->] (.4,-.2) -- (-.2,-.3);
   \end{scope}
\end{tikzpicture}
  \caption{The statistical signed density $\lambda_n^+ - \lambda_n^-$ (recall Figure \ref{fig:rhopm:stat:no-ann}) compared to $\rho^+ - \rho^-$ and $\kappa$. }
     \label{fig:kap:stat:no-ann}
\end{figure}
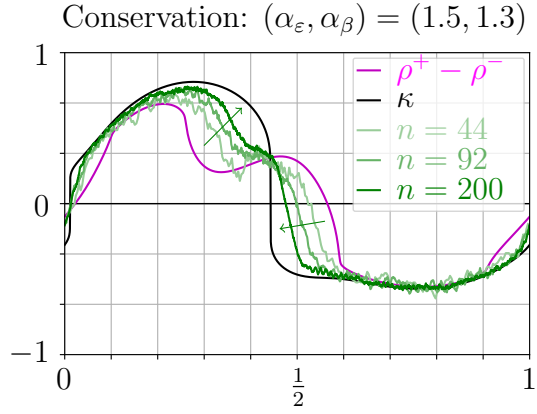

\begin{figure}[h!]
   \centering
   \begin{tikzpicture}
    \draw (0,-.3) node{\includegraphics[width = .4\textwidth]{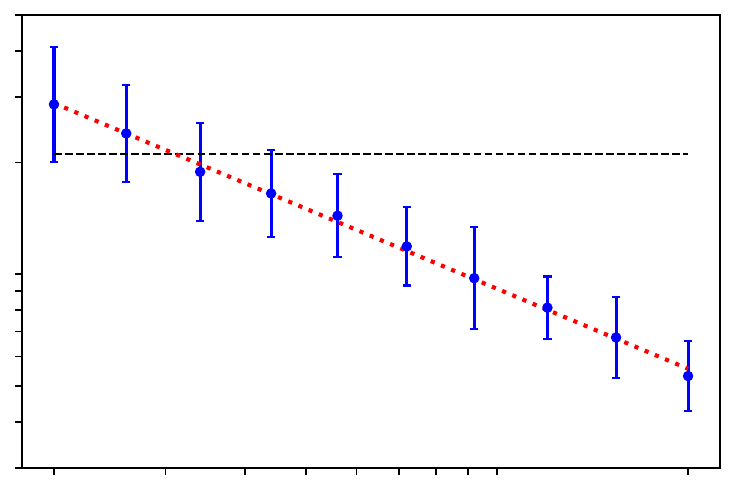}}; 
   \draw[red] (-1.3, .35) -- (-1.3, 0) -- (-.5, 0); 
   \draw[red, ->] (-1.1, -1.5) node[below] {$\sim n^{-0.72}$} -- (-1.1, -.1);
   \draw (0,-2.8) node[below]{Number of dislocations, $n$};
   \draw (1.15,-2.3) node[below]{$100$};
   \draw (2.8,-2.3) node[below]{$200$};
   \draw (-2.8,-2.3) node[below]{$20$};
   \draw (0,1.9) node[above]{$\disccon$ $\to$ \eqref{Pann}: \ $(\alpha_\e, \alpha_\beta) = (1.5, 1.3)$ };
   \draw (-4.5,-.3) node[rotate=90]{Distance, $w$};
   \draw (2, .5) node[above] {$w_\infty$};
   \draw (-3.1,-2.25) node[left]{$0.003$};
   \draw (-3.1,-0.55) node[left]{$0.01$};
   \draw (-3.1, 1.75) node[left]{$0.05$};
 \end{tikzpicture}
   \caption{The data is plotted similar to Figure \ref{fig:W:ref:val}, but now for the data point $(\alpha_\e, \alpha_\beta) = (1.5, 1.3)$, the discrete solution of $\disccon$ and the continuous solution from \eqref{Pann}. }
      \label{fig:W:no-ann:kap}
 \end{figure}
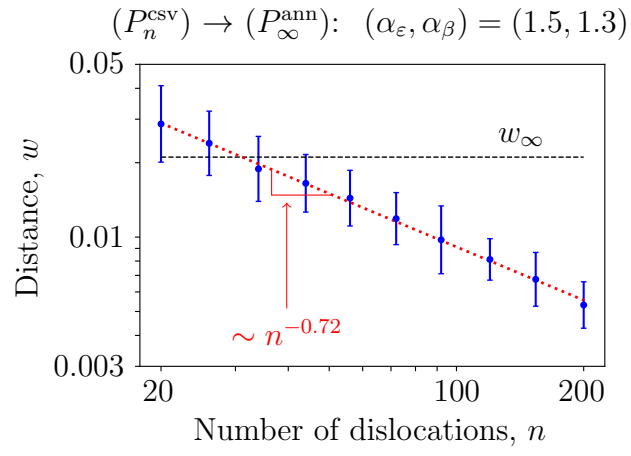

A growing accumulation is reminiscent of the \textit{annihilation} model, where all annihilation happens at  single points; the fronts of $\kappa$. To test whether the statistical density tends instead towards the continuous solution $\kappa$ of \eqref{Pann}, we plot in Figure~\ref{fig:kap:stat:no-ann} $\lambda_n^+ - \lambda_n^-$ and $\kappa$. Indeed, the plots show that, as $n$ increases, $\lambda_n^+ - \lambda_n^-$ moves away from $\rho^+ - \rho^-$ and towards $\kappa$. To test whether not just the statistical density, but actually each sample path, converges, we plot the corresponding $w$ (i.e.\ with input from $\disccon$ and \eqref{Pann}) in Figure \ref{fig:W:no-ann:kap}. The results are striking; $w$ converges just as well as in Figure \ref{fig:W:ref:val} for the reference values.

\clearpage

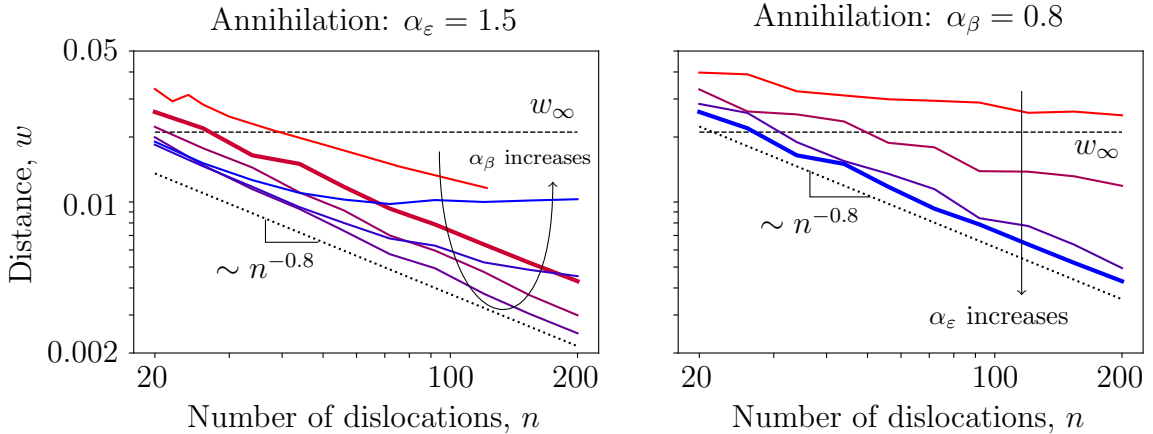
\begin{figure}[h!]
  \centering
  \begin{tikzpicture}
   \def\r {.15}      
      
   \draw (0,-.3) node{\includegraphics[width = .4\textwidth]{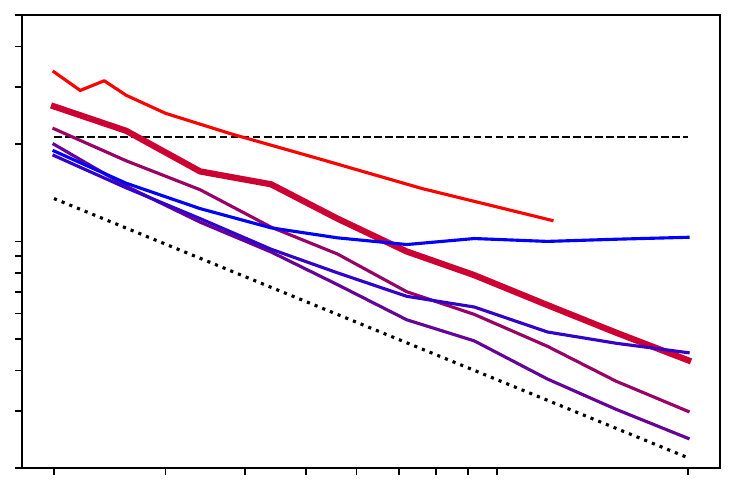}}; 
   \draw (-1.3, -.5) -- (-1.3, -.8) node[below] {$\sim n^{-0.8}$} -- (-.6, -.8); 
   \draw (0,-2.8) node[below]{Number of dislocations, $n$};
   \draw (1.15,-2.3) node[below]{$100$};
   \draw (2.8,-2.3) node[below]{$200$};
   \draw (-2.8,-2.3) node[below]{$20$};
   \draw (0,1.8) node[above]{Annihilation: $\alpha_\e = 1.5$};
   \draw (-4.5,-.3) node[rotate=90]{Distance, $w$};
   \draw (2.5, .7) node[above] {$w_\infty$};
   \draw (-3.1,-2.25) node[left]{$0.002$};
   \draw (-3.1,-0.3) node[left]{$0.01$};
   \draw (-3.1, 1.75) node[left]{$0.05$};

  \draw[->] (1,.4) .. controls (1,-2.1) and (2.5,-2.5) .. (2.5,0);
  \draw (2.2,0) node[above]{\scriptsize $\alpha_\beta$ increases};

   \begin{scope}[shift={(7.2, 0)}]  
     \draw (0,-.3) node{\includegraphics[width = .4\textwidth]{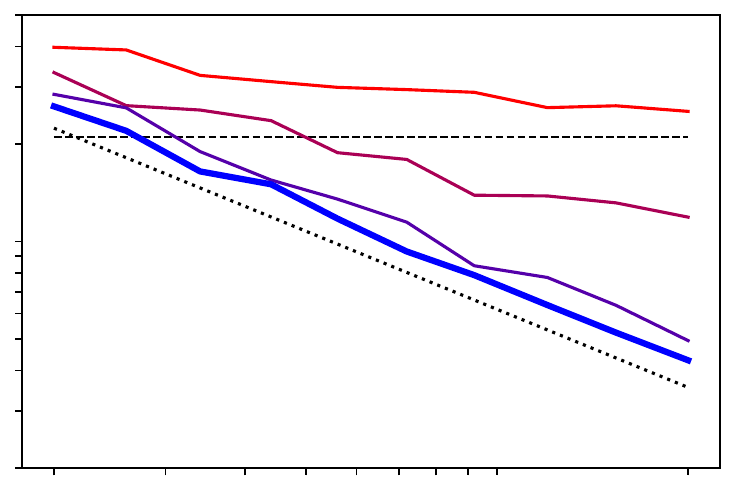}};
   \draw (-1.3, .15) -- (-1.3, -.2) node[below] {$\sim n^{-0.8}$} -- (-.5, -.2); 
   \draw (1.15,-2.3) node[below]{$100$};
   \draw (2.8,-2.3) node[below]{$200$};
   \draw (-2.8,-2.3) node[below]{$20$};
   \draw (0,-2.8) node[below]{Number of dislocations, $n$};
   \draw (0,1.8) node[above]{Annihilation: $\alpha_\beta = 0.8$};
   \draw (2.5, .7) node[below] {$w_\infty$};
   \draw[->] (1.5,1.2) -- (1.5,-1.5);
   \draw (1.2,-1.5) node[below]{\footnotesize $\alpha_\e$ increases};
   \end{scope}
\end{tikzpicture}
\caption{Plots of the median distances between discrete and continuum models with annihilation as $n$ varies for all choices of the exponents $(\alpha_\e, \alpha_\beta)$ in Figure~\ref{fig:epsbeta}. Left: exponent $\alpha_\e=1.5$ is fixed, and plots for varying $\alpha_\beta = 0.5, 0.8, 1.0, 1.3, 1.4, 1.5$ are shown on the colour scale from red to blue. The curved arrow illustrates the trend for increasing $\alpha_\beta$. Right: $\alpha_\beta=0.8$ is fixed, and results for $\alpha_\e = 1.0, 1.1, 1.2, 1.5$ are plotted with the same red-to-blue colour scheme. The thick graph in both plots corresponds to the reference values $(\alpha_\e, \alpha_\beta) = (1.5, 0.8)$, and represents the same data shown on the right of Figure~\ref{fig:W:ref:val}.  }
     \label{fig:W:ann}
\end{figure}

\paragraph{Results for the annihilation models.}
Next, we investigate the hypothesis that $\discann$ $\to \contann$. In contrast to the conservation models, Figure~\ref{fig:W:ann} strongly suggests that this convergence holds for most of the exponents $(\alpha_\e, \alpha_\beta)$ investigated. We find that the rate of convergence for the median distance between discrete and continuum densities is roughly $w\sim n^{-0.8}$ for all such data points.
Only the exponent pairs close to the boundary of the region in Figure~\ref{fig:epsbeta} exhibit different behaviour: 
\begin{itemize}\itemsep0em
  \item As $\alpha_\e$ approaches $1$ with $\alpha_\beta=0.8$ fixed, it appears that the rate of convergence decreases. At $\alpha_\e = 1$, i.e.\ on the boundary of our parameter regime, we do not expect convergence from a theoretical viewpoint (recall Section \ref{sec:models:parameters}), and we indeed observe here that $w$ remains above $w_\infty$. 
  \item As $\alpha_\beta$ approaches $1.5$, it appears that $w$ ceases to decrease in $n$ for $n$ sufficiently large.  
  \item As  $\alpha_\beta$ approaches $0$, our results appear inconclusive: we did not obtain enough data to get good predictions for $w$. The reason is the computational cost, which turns out to be larger than predicted in \eqref{comp:time}. This might be due to many fast transitions which do not advance simulation time much. 
\end{itemize}

\subsection{Discussion}
Our numerical simulations provide strong evidence that the models with annihilation $\discann$ converge to $\contann$ in the asymptotic regime studied, i.e. where $1\ll n,\beta\ll \frac{1}{\e}$, as outlined in Section~\ref{sec:models:parameters}. As we approach the boundaries of this regime, this convergence weakens and appears to fail. This failure is expected as the PDE model $\contann$ is derived on the basis of the particular asymptotic ordering, but convergence appears consistent away from these boundary cases.

In contrast, there is little evidence for the convergence of the conservative models $\disccon$ to $\contcon$. In fact, we obtain evidence that $\disccon$ does not converge to $\contcon$ in some of the simulations within the asymptotic parameter regime studied. Hence, the asymptotic regime where the convergence of $\disccon$ to $\contcon$ might be conjectured is significantly smaller than the regime $1\ll n,\beta\ll \frac{1}{\e}$. In contrast, we find evidence that in fact $\disccon$ may be better approximated by the PDE model with annihilation, $\contann$. One possible explanation for this observation is that during the evolution of $\disccon$, dipoles where dislocations of opposite sign collide behave as if effectively removed. Even if they are not actually removed, the probability of them dissociating at later times is very small, and they do not impede the motion of other dislocations in the system. Hence, $\contann$ may provide a better macroscopic model. As seen in Figure~\ref{fig:rho:kap:intro}, there are genuine differences in the evolution of the net Burgers vector in regions where dislocations meet, and so the distinction between the PDE models is significant. We note that this particular explanation is in line with the counter-example to convergence in \cite{GarroniVanMeursPeletierScardia19} for a related ODE model. 

\section{Conclusion}
\label{sec:conclusion}

In this work, we have provided a careful numerical study of the asymptotic behaviour for two models of dislocation dynamics where dislocation annihilation is treated differently. We obtained strong numerical evidence for the convergence of the discrete model where annihilation is explicitly treated to the corresponding PDE model in the parameter regime where the number of particles and the inverse temperature are large, but not as large as the inverse lattice spacing. In contrast, we obtained evidence that a density-conserving PDE model does not provide an accurate description of the dynamics in the same parameter regime, even when the microscopic model mirrors the assumption that dislocation dipoles are free to separate at later time. 

It should be noted that our results have been obtained for a one-dimensional dislocation model in which only a single slip-plane is active: situations in which multiple slip planes under different angles are active and in which dislocation climb is possible may yield different results. Nevertheless, already in our one-dimensional setup, our conclusions reveal that the treatment of dislocation collision in discrete and continuum models is an important aspect of modelling that requires further mathematical investigation: we have seen that different assumptions can yield qualitatively different evolution features, which has consequences for the accuracy of PDE models of dislocations. Hence, we conjecture that dislocation collisions are of significant importance in two- and three-dimensional dislocation modelling, and therefore for the modelling of macroscopic plastic behaviour in metals.

\section*{Acknowledgements}

T.\ Hudson is grateful for partial support from UKRI during this work through NSF-UKRI grant reference UKRI3611.
P.\ van Meurs has received financial support from JSPS KAKENHI Grant Number JP24K06843.

\newcommand{\etalchar}[1]{$^{#1}$}

\end{document}